\documentclass[]{JHEP3} 

\usepackage{epsfig,multicol,bbm}
\usepackage[sort&compress,numbers]{natbib}

\newcommand\fverb{\setbox\fverbbox=\hbox\bgroup\verb}
\newcommand\fverbdo{\egroup\medskip\noindent%
			\fbox{\unhbox\fverbbox}\ }
\newcommand\fverbit{\egroup\item[\fbox{\unhbox\fverbbox}]}
\newbox\fverbbox

\usepackage{graphicx}
\def\nf{n_{\mskip-2mu f}}
\def\prop#1{{\cal P}_{#1}}

\def\Ls{\mathrm{Ls}}
\def\Li{\mathrm{Li}}
\def\Ll{\mathrm{L}}
\def\I33m{\mathrm{I}_3^{3{\mathrm m}}}
\def\nn{\nonumber}

\def\be{\begin{equation}}
\def\ee{\end{equation}}
\def\bea{\begin{eqnarray}}
\def\eea{\end{eqnarray}}

\def\qb{{\bar{q}}}

\def\spa#1.#2{\left\langle#1\,#2\right\rangle}
\def\spb#1.#2{\left[#1\,#2\right]}
\def\lor#1.#2{\left(#1\,#2\right)}
\def\sand#1.#2.#3{%
\left\langle\smash{#1}{\vphantom1}^{-}\right|{#2}%
\left|\smash{#3}{\vphantom1}^{-}\right\rangle}
\def\sandp#1.#2.#3{%
\left\langle\smash{#1}{\vphantom1}^{-}\right|{#2}%
\left|\smash{#3}{\vphantom1}^{+}\right\rangle}
\def\sandpp#1.#2.#3{%
\left\langle\smash{#1}{\vphantom1}^{+}\right|{#2}%
\left|\smash{#3}{\vphantom1}^{+}\right\rangle}
\def\sandpm#1.#2.#3{%
\left\langle\smash{#1}{\vphantom1}^{+}\right|{#2}%
\left|\smash{#3}{\vphantom1}^{-}\right\rangle}
\def\sandmp#1.#2.#3{%
\left\langle\smash{#1}{\vphantom1}^{-}\right|{#2}%
\left|\smash{#3}{\vphantom1}^{+}\right\rangle}
\def\spab#1.#2.#3{\langle#1|#2|#3]}
\def\spba#1.#2.#3{[#1|#2|#3\rangle}
\def\Etmax{E_T^{\rm{max}}}
\def\fig{Fig.}
\def\tab{Table}

\title{Vector boson pair production at the LHC}

\author{
    John M. Campbell, R. Keith Ellis and Ciaran Williams
    \\
    Fermilab, Batavia, IL 60510, USA
    \\
    E-mails: 
    {\tt johnmc@fnal.gov}, 
    {\tt ellis@fnal.gov}, 
    {\tt ciaran@fnal.gov}.}

\received{\today} 		
\accepted{\today}		

\preprint{
FERMILAB-PUB-11-182-T}

\abstract{
We present phenomenological results for vector boson pair production at the LHC, obtained using the parton-level
next-to-leading order program MCFM. We include the  implementation of a new process in the code,
$pp\rightarrow\gamma\gamma$, and important updates to existing processes. We incorporate fragmentation contributions in
order to allow for the experimental isolation of photons in $\gamma\gamma$, $W\gamma$, and $Z\gamma$ production and also
account for gluon--gluon initial state contributions for  all relevant processes. We present results for a variety of
phenomenological scenarios, at the current operating energy of $\sqrt{s} = 7$~TeV and for the ultimate machine goal,
$\sqrt{s} = 14$~TeV. We investigate the impact of our predictions on several important distributions that
enter into searches for new physics at the LHC.}


\begin{document} 



\section{Introduction}
The current plan for the LHC calls for running 
in both 2011 and 2012. Running in 2011 is at a centre of mass energy
at $\sqrt{s}=7$~TeV, with a baseline expectation of $1\,\mbox{fb}^{-1}$ per 
experiment and a good chance that greater luminosity will be accumulated.
At the end of the 2012 run it is likely that data samples in excess
of $5\,\mbox{fb}^{-1}$ will have been accumulated by both of the general purpose
detectors. Data samples of this size will (at the very least) allow
detailed studies of the production of pairs of vector bosons.

\begin{figure}
\begin{center}
\includegraphics[width=12cm]{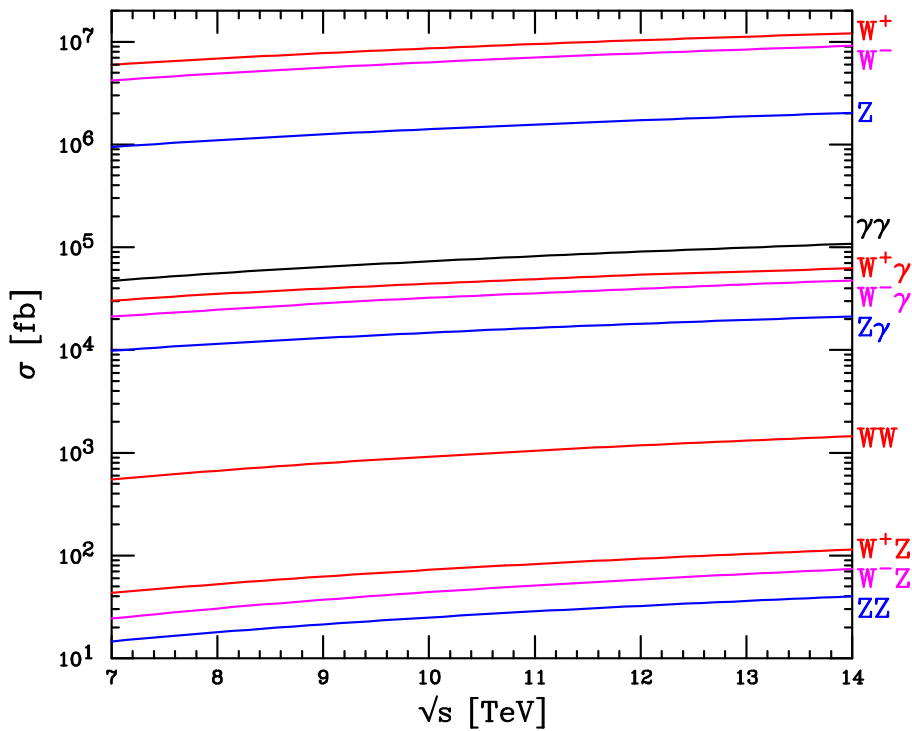}
\end{center}
\caption{NLO boson production in $pp$-collisions. The decay branching ratios
of the $W$'s and $Z$'s into one species of leptons are included.
For $\gamma\gamma$ and $V\gamma$ we apply $p_{T}$ cuts of 25 and 10 GeV to photons respectively.}
\label{DiB:BR}
\end{figure}

It therefore seems opportune to provide up-to-date
predictions for the production of all pairs of vector bosons,
specifically for the LHC operating at $7$~TeV. This extends the previous implementation of diboson 
production in MCFM~\cite{Campbell:1999ah} which was focussed primarily on the Tevatron. Moreover, we also
consider the production of final states that contain real photons. This requires the inclusion of
fragmentation contributions in order to address the issue of isolation in an experimental context. In
addition, we have also included the contribution of the gluon--gluon
initial state to a number of processes. These finite corrections are
formally of higher order but can be of phenomenological relevance at
the LHC where the gluon flux is substantial. 

A review of the current experimental status of vector pair boson
production, primarily from the Tevatron, can be found in
ref.~\cite{Hobbs:2010yg}.  The production of pairs of vector bosons is
crucial both in order to check the gauge structure of the Standard
Model (SM) and in the search for new physics.  This is because
production of vector boson pairs and the associated particles from
their decay, enter as irreducible backgrounds for many Higgs and
new physics searches.  The observationally most promising decays of
the Higgs boson are to two photons (for a light Higgs), or to two $W$'s
or two $Z$'s that decay leptonically.  Clearly vector boson pair
production is an irreducible background in these searches. Processes
with leptons and missing energy are typical signatures of many new
physics models, of which supersymmetry is a classic example.  Again,
knowledge of SM processes which possess multiple leptons and missing
energy is crucial in the quest to discover or rule out these models.

In \fig~\ref{DiB:BR} we show the rates for various electroweak
processes at energies between $\sqrt{s}=7$ and $14$~TeV.  This figure
serves both as a road-map to this paper and as an indication of the
relative size of the various diboson processes. We present the cross
sections for single boson production to illustrate the orders of
magnitude which separate single boson and diboson production.  Where
appropriate we have included the branching ratios of vector bosons to
a single family of leptons and applied a transverse momentum cut of
10~GeV ($W\gamma$ and $Z\gamma$) and 25~GeV $(\gamma\gamma)$ to
photons. No other cuts are applied to the boson decay products.

Updating the diboson processes in MCFM for the new energy range
probed at the LHC is the primary aim of this work. With that in mind
we begin in section~\ref{sec:frag} by outlining the steps needed to
include photon fragmentation in the code. Section~\ref{sec:overview}
serves as an overview, describing the parameters that we use and
outlining the processes that receive extra corrections from gluon
initiated production mechanisms.  Section~\ref{sec:gamgam} discusses
the phenomenology of $\gamma\gamma$ production at the LHC. We
investigate the role of isolation on the cross section and the impact
of Higgs search cuts on di-photon production. Sections~\ref{sec:Wgam}
and~\ref{sec:Zgam} contain our predictions for $W\gamma$ and $Z\gamma$
production at the LHC. We investigate the role of final-state
radiation in our calculations and compare our NLO results with the
recently reported cross sections from
CMS~\cite{EWK-10-008}. Sections~\ref{sec:WW},~\ref{sec:WZ}
and~\ref{sec:ZZ} turn to the production of two massive vector bosons.
We are able to compare our prediction for the $WW$ cross section with
early results from ATLAS and
CMS~\cite{Collaboration:2011kk,Chatrchyan:2011tz}.  We examine the effect
of the gluon initiated processes in the $WW$ and $ZZ$ final states,
with particular emphasis on their role as Higgs backgrounds. For $WZ$
production we discuss briefly the properties of boosted $Z$'s. Finally
in section~\ref{sec:conc} we draw our
conclusions. Appendix~\ref{app:inputs} contains a more detailed
discussion of our electroweak parameters whilst
appendix~\ref{app:amps} presents formulae for the $gg \rightarrow
V_1V_2$ amplitudes as implemented in MCFM.


\section{Photon fragmentation}
\label{sec:frag}
Since we will consider a number of final states including photons we must first discuss the additional
complications that this involves, compared to the production of $W$ and $Z$ bosons.
Experimentally, the production of photons occurs via two mechanisms. Prompt photons are produced in hard scattering
processes  whilst secondary photons arise from the decays of particles such as the $\pi^{0}$. Since secondary
photons are typically associated  with hadronic activity one can attempt to separate these contributions by
limiting the amount of hadronic energy in a cone of size $R_0=\surd(\Delta \eta^2 +\Delta \phi^2) $ around the photon. 
Experimental isolation cuts are of the form, 
\begin{eqnarray}
\sum_{\in R_0} E_T{\rm{(had)}} < \epsilon_h \, p^{\gamma}_{T} \qquad {\rm{or}}
 \qquad \sum_{\in R_0} E_T{\rm{(had)}} < \Etmax \;.
\label{eq:iso}
\end{eqnarray}
Thus the transverse hadronic energy, $E_T{\rm{(had)}}$, is limited to be some small fraction of the
transverse momentum of the photon or cut off at a fixed, small upper limit. 

Matters are complicated both experimentally and theoretically by a second source of prompt photons. A hard QCD
parton can fragment non-perturbatively into a photon. As a result a typical photon production cross section
takes the form,
\begin{eqnarray}
\sigma = \sigma^{\gamma}(M^2_{F}) + \int dz \, D^{a}(z) \sigma^{a}(z,M^2_F). 
\end{eqnarray}
Here $\sigma^{\gamma}$ represents the direct component of the photon
production cross section whilst the second term arises from the
fragmentation of a parton $a$ into a photon with momentum $z p^{a}$.
Each contribution separately depends on the fragmentation scale,
$M_F$.  The fragmentation functions, taken as solutions to a DGLAP
equation are of (leading) order $\alpha_{EW}/\alpha_{s}$. This means
that they are formally of the same order as the leading order direct
term. At high-energy hadron colliders, the QCD tree-level matrix
element, coupled to a fragmentation function can become the dominant
source of prompt photon production. However, the magnitude of these
terms can be drastically reduced by applying the isolation cuts
described above. This is due to the fact that the fragmentation
functions strongly favour the low $z$ region. Once the photon is
isolated, $z$ is typically large enough that the fragmentation
contribution drops substantially from the unisolated case.

A theoretical description of isolated photons is complicated because of the occurence of collinear singularities
between photons and final-state quarks.  A finite cross section is only obtained when these singularities are absorbed
into the fragmentation functions.  As a result the only theoretically well-defined NLO quantity is the sum of the direct
and fragmentation contributions.
Once these two contributions are included one can isolate the photon using the cuts of Eq.~(\ref{eq:iso}) in an
infrared safe way~\cite{Catani:2002ny}. 

Although the underlying dynamics of photon fragmentation are non-perturbative the evolution of the functions with
the scale $M_F$ is perturbative. In the same manner as the parton distribution functions, the fragmentation functions
satisfy a DGLAP evolution equation. In MCFM we use the fragmentation functions of ref.~\cite{Bourhis:1997yu}, which are NLL solutions to the DGLAP
equation. 

Final state quark-photon collinear singularities are removed using a variant~\cite{Catani:2002ny} of the
Catani-Seymour dipole subtraction formalism~\cite{Catani:1996vz}. More specifically, we treat the photon in the same
manner as one would  treat an identified final state parton (with the appropriate change of colour and coupling factors). 
Integration of these subtraction terms over the additional parton phase space yields pole pieces of the form~\cite{Catani:2002ny},
\begin{eqnarray}
D^{\gamma}_{q} = -\frac{1}{\epsilon}\frac{\Gamma(1-\epsilon)}{\Gamma(1-2\epsilon)}\bigg(\frac{4\pi\mu^2}{M_F^2}\bigg)\frac{\alpha}{2\pi} \, e^2_q P_{\gamma q}(z)
 \;, 
\end{eqnarray}
where $P_{\gamma q}(z)$ is the tree level photon-quark splitting function. This piece $D^{\gamma}_q$ is the lowest order definition of the photon fragmentation 
function in the $\overline{MS}$ scheme. This singularity is then absorbed into the fragmentation functions to yield finite cross sections.

Since the isolation cuts reduce the magnitude of the fragmentation contributions we calculate the QCD matrix elements
$\sigma^{a}(z,M^2_F)$ to LO, i.e. we neglect  NLO corrections to the fragmentation processes.

An alternative procedure, in which one can avoid calculating the fragmentation contributions altogether,
is to follow the smooth cone isolation of Frixione~\cite{Frixione:1998jh}.
In such an approach one applies the following isolation prescription to the photon, 
\begin{eqnarray}
\sum_{R_{j\gamma} \in R_0} E_T{\rm{(had)}} < \epsilon_h p_{T}^{\gamma} \bigg(\frac{1-\cos{R_{j\gamma}}}{1-\cos{R_0}}\bigg).
\end{eqnarray}
Using this prescription, soft radiation is allowed inside the photon cone but collinear singularities are removed. Since the smooth-cone isolation
is infra-red finite, there is no need to include fragmentation contributions in this case.
Currently this isolation is difficult to implement experimentally and
therefore it is not used in this paper.~\footnote{Smooth cone isolation is however available in MCFM for theoretical comparisons.}


\section{Overview}
\label{sec:overview}

The results presented in this paper are obtained with the latest version of
the MCFM code (v6.0). We use the default set of electroweak parameters as
described in Appendix~\ref{app:inputs}.

For the parton distribution functions (pdfs) we use the sets of Martin, Stirling,
Thorne and Watt~\cite{Martin:2009iq}. For the calculation of the LO results presented here we employ
the corresponding LO pdf fit, with 1-loop running of the strong coupling and
$\alpha_s(M_Z)=0.13939$. Similarly, at NLO we use the NLO pdf fit, with
$\alpha_s(M_Z)=0.12018$ and 2-loop running. The fragmentation of partons into
photons uses the parametrization ``set II'' of Bourhis, Fontannaz and
Guillet~\cite{Bourhis:1997yu}. 

As mentioned in the introduction, for several processes we have included contributions of the form  $gg\rightarrow
V_1V_2$. These contributions proceed through a closed fermion loop and form a gauge  invariant subset of the one-loop
amplitudes. However, since there is no $gg$ tree level contribution the first  time these pieces enter in the
perturbative expansion is at $\alpha_S^2$, (i.e.\ NNLO). Simple power counting would  thus lead one to assume that these pieces
are small, of the order of a few percent of the LO cross section. At the LHC this is often  not the case, since the
large gluon flux in the pdfs can overcome the ${\cal O}(\alpha_S^2)$ suppression in the perturbative expansion. The
resulting gluon-gluon contributions are instead ${\cal O}(10\%)$  of the LO cross section, i.e. these pieces are
comparable to the other NLO contributions. 

Charge conservation ensures that not all diboson processes receive these gluon-gluon initiated contributions.
The allowed processes are
$gg\rightarrow \{\gamma\gamma, Z\gamma,W^+W^-,ZZ\}$, each of which has been studied in some format in the
past~\cite{Dicus:1987fk,Nadolsky:2007ba,Bern:2001df,Bern:2002jx,Ametller:1985di,vanderBij:1988fb,Adamson:2002rm,Dicus:1987dj,Glover:1988fe,Glover:1988rg,Binoth:2005ua,Binoth:2006mf,Matsuura:1991pj,Zecher:1994kb,:2008uu,Binoth:2008pr}.
We refer the reader to the appropriate section for details of each calculation. We note that for $gg\rightarrow
\{Z\gamma,W^+W^-,ZZ\}\rightarrow$ leptons, we present (to the best of our knowledge) analytic formulae for the
helicity amplitudes for the first time. These formulae were readily obtained from the amplitudes for
the process $e^+e^-\rightarrow$ 4 partons~\cite{Bern:1997sc}. 

Since there are no $gg$ tree level contributions, each of these 1-loop amplitudes is both infrared and ultraviolet
finite. This means that, once calculated, these contributions are simple to implement in MCFM. Throughout this paper we
include these pieces in the NLO results {\it{except}} for the $gg\rightarrow\gamma\gamma$ section. 
Since the strong corrections to the $gg\rightarrow\gamma\gamma$ process, including the two-loop
amplitude~\cite{Bern:2001df} are known and sizeable~\cite{Bern:2002jx}, we proceed differently for this process. 
The two-loop $gg\rightarrow\gamma\gamma$ amplitude is infrared divergent and must be combined with real radiative
corrections in exactly the same manner as a canonical NLO contribution~\cite{Bern:2002jx}.
As a result these contributions are included in our NLO predictions for the diphoton process, while the one-loop
$gg\rightarrow\gamma\gamma$ calculation is included in the LO result.


\section{$\gamma\gamma$ production}
\label{sec:gamgam}

\subsection{Description of the calculation}
In view of its role as the principal background in the search 
for the light Higgs boson in the decay mode $H \to \gamma \gamma$,
it is important that the prediction for Standard Model diphoton
production is as accurate as possible.
The production of photons in hadron-hadron interactions proceeds
through the Born level process,
\begin{equation}
q + \bar{q} \to \gamma \gamma \;.
\end{equation}
Corrections to this picture due to QCD interactions have been first
considered at $O(\alpha_s)$ in 
ref.~\cite{Aurenche:1985yk} and the results for that process
have been included in the Diphox Monte Carlo~\cite{Binoth:1999qq}.
The large flux of gluons at high energy -- in particular at current LHC
energies -- means that diagrams involving loops of quarks can give a significant
additional contribution~\cite{Ametller:1985di,Dicus:1987fk,Nadolsky:2007ba},
\begin{equation}
g + g \to \gamma \gamma \;.
\end{equation}
Since these contributions can be rather large, in order to obtain a reliable estimate 
of their contribution to the diphoton cross section it is necessary to include higher order
corrections. The results of such a calculation, involving two-loop virtual
contributions~\cite{Bern:2001df}, were presented in ref.~\cite{Bern:2002jx}.

The results presented in this section are obtained using our current implementation in MCFM
which is as follows. The $gg$ process is included at NLO using the two loop matrix elements
of ref.~\cite{Bern:2001df} and following the implementation of ref.~\cite{Bern:2002jx}.
We include five flavours of massless quarks and neglect the effect of the top quark loops,
which are suppressed by $1/m_t^4$.
Next-to-leading order corrections to the $q{\bar q}$ initiated process are more
straightforward to include, although some care is required due to the issues of
photon fragmentation and isolation that have been described in section~\ref{sec:frag}.

We can compare our implementation of $pp\rightarrow \gamma\gamma$ to Diphox~\cite{Binoth:1999qq}. 
Diphox contains NLO predictions for both the direct and fragmentation pieces, but only includes the $gg$ initiated 
pieces at leading order. In MCFM we include NLO predictions for the direct pieces, LO predictions for the fragmentation processes (using NLL 
fragmentation functions) and the ``NLO'' $gg$ predictions. For isolated photons the ``NLO'' $gg$ corrections represent
around $5\%$ of the total cross section, so we expect them to be at least as important as the NLO corrections to the fragmentation piece.

\subsection{Results}

As a point of reference, we first consider the cross section for 
unisolated photons at the LHC, for various centre-of-mass energies.
We apply only basic acceptance cuts on the two photons,
\begin{eqnarray}
p_{T}^{\gamma} > 25~{\rm{GeV}} \;, \quad |\eta_{\gamma}| < 5 \;.
\label{eq:gamgambasiccuts}
\end{eqnarray} 
The cross sections we report are completely inclusive in any additional parton radiation. For our theoretical predictions
we choose renormalisation ($\mu_R$), factorisation ($\mu_F$) and fragmentation scales ($M_F$)
all equal to the diphoton invariant mass, $m_{\gamma \gamma}$.
The results of our study at LO and NLO are shown in \tab~\ref{gamgamtot}, where the
percentage uncertainties quoted on the NLO cross sections are estimated by varying all
scales simultaneously by a factor of two in each direction.
 \renewcommand{\baselinestretch}{1.6}
 \begin{table}[]
 \begin{center}
 \begin{tabular}{|c|c|c|}
 \hline
 $\sqrt{s}$~[TeV] & $\sigma^{LO}(\gamma\gamma) $~[pb] &  $\sigma^{NLO}(\gamma\gamma) $~[pb]\\
 \hline
    7 &  35.98(0)&  47.0(1)$^{+  5 \%}_{ -6\%}$ \\
    8 &  43.04(1)&  55.8(1)$^{+  4 \%}_{ -6\%}$ \\
    9 &  50.32(1)&  64.3(1)$^{+  5 \%}_{ -5\%}$ \\
   10 &  57.76(1)&  73.0(2)$^{+  4 \%}_{ -5\%}$ \\
   11 &  65.37(1)&  81.8(2)$^{+  3 \%}_{ -5\%}$ \\
   12 &  73.07(1)&  90.5(3)$^{+  4 \%}_{ -5\%}$ \\
   13 &  80.89(1)&  99.1(3)$^{+  4 \%}_{ -5\%}$ \\
   14 &  88.76(2)& 108.1(3)$^{+  3 \%}_{ -5\%}$ \\
 \hline
 \end{tabular}
  \renewcommand{\baselinestretch}{1.0}
 \caption{LO and NLO cross sections for diphoton production at the LHC with the acceptance
 cuts of Eq.~(\protect\ref{eq:gamgambasiccuts}), as a function of
 $\sqrt{s}$. The Monte Carlo integration error on each prediction is shown in parentheses.
 For the NLO results the theoretical scale uncertainty is computed according to the procedure
 described in the text and is shown as a percentage deviation.}
 \label{gamgamtot}
 \end{center}
 \end{table}
  \renewcommand{\baselinestretch}{1.0}
The inclusion of both $gg$ and $q\bar q$ processes in the LO result, and the next order
corrections to both at NLO, results in only a mild $20$--$30$\% increase in the cross
section at NLO. Moreover, these predictions are rather stable with respect to scale variations
over the range studied, with deviations in each direction of at most $6$\%.

\begin{figure} 
\begin{center}
\includegraphics[width=12cm]{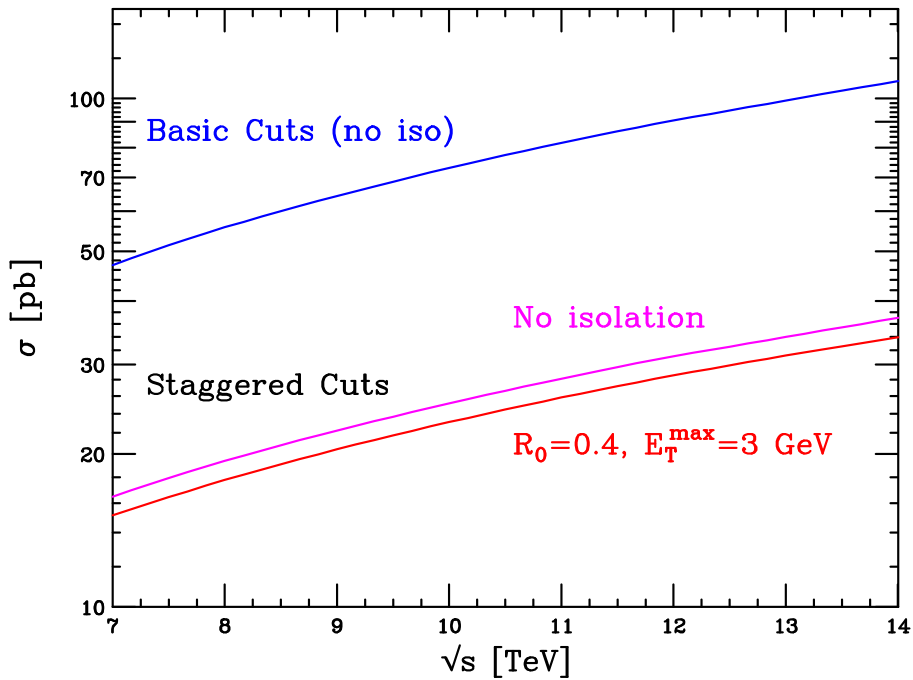}
\caption{The NLO prediction for the diphoton cross section (in picobarns) as
a function of the centre of mass energy, $\sqrt{s}$. The cross sections are shown
for three sets of cuts: only the basic cuts of Eq.~(\protect\ref{eq:gamgambasiccuts}) (upper, blue
curve); the staggered cuts of Eq.~(\protect\ref{eq:gamgamstagcuts}) (middle, magenta curve);
the isolated photon cross section, Eqs.~(\protect\ref{eq:gamgamstagcuts},~\protect\ref{eq:gamgamisol})
(lower, red curve).
\label{fig:cross}}
\end{center}
\end{figure}

We now wish to investigate a more realistic set of cuts in which the photon is isolated.
Since this final state is particularly interesting in the context of a low-mass Higgs
search~\cite{:2008uu}, for illustration we adopt the set of cuts used in an early search
by the ATLAS collaboration~\cite{ATLAS-CONF-2011-025}.
The photons are required to be relatively central and subject to staggered transverse
momentum cuts,
\begin{equation}
p_{T}^{\gamma_1} > 40~{\rm{GeV}} \;, \quad p_{T}^{\gamma_2} > 25~{\rm{GeV}} \;,
 \quad |\eta_{\gamma_i}| < 2.5 \;,
\label{eq:gamgamstagcuts}
\end{equation} 
and are isolated using a fixed maximum hadronic energy in a photon cone (c.f. Eq.~(\ref{eq:iso})),
\begin{equation}
 \quad R_0 = 0.4 \;, \quad \Etmax=3~\rm{GeV} \;.
\label{eq:gamgamisol}
\end{equation} 
The effect of these cuts, as a function of $\sqrt{s}$, is shown in \fig~\ref{fig:cross}.
The effect of the staggered cuts, Eq.~(\ref{eq:gamgamstagcuts}), is to lower the cross
section by approximately a factor of three compared to the basic cuts of
Eq.~(\ref{eq:gamgambasiccuts}). The isolation condition, Eq.~(\ref{eq:gamgamisol}), further
reduces the cross section from the nominal unisolated prediction by about 9\%. We note that 
this reduction is smaller than one would typically expect when going from unisolated to isolated 
cross sections. This is due mostly to the staggered cuts which favour the 3 particle final state.

\begin{figure} 
\begin{center}
\includegraphics[width=12cm]{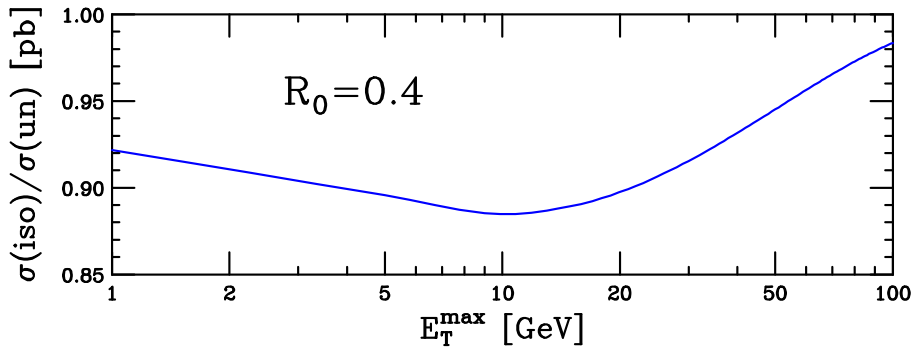}
\caption{The fraction of the unisolated diphoton cross section that remains when the photon is isolated,
as a function of the maximum amount of transverse hadronic energy allowed in the photon isolation cone, $\Etmax$.
The centre of mass energy is $\sqrt{s}=7$~TeV and photons are identified according to the staggered cuts
of Eq.~(\protect\ref{eq:gamgamstagcuts}). The radius of the isolation cone is $R_0=0.4$. 
\label{fig:E_H}}
\end{center}
\end{figure}

In fact the cross section is rather insensitive to the amount of transverse
hadronic energy allowed in the isolation cone. This is illustrated in \fig~\ref{fig:E_H},
which shows the dependence  of the cross section on the value of the isolation parameter
$\Etmax$.
As a result of the small variation over this range, isolation cuts of the form $E+\delta
p^{\gamma}_{T}$ where $E$ and $\delta$ are constants  and $\delta \ll E$ are well-approximated
theoretically by using a simple constant $\Etmax = E+\delta p^{\gamma}_{T,\rm{min}}$.

\begin{figure} 
\begin{center}
\includegraphics[width=12cm]{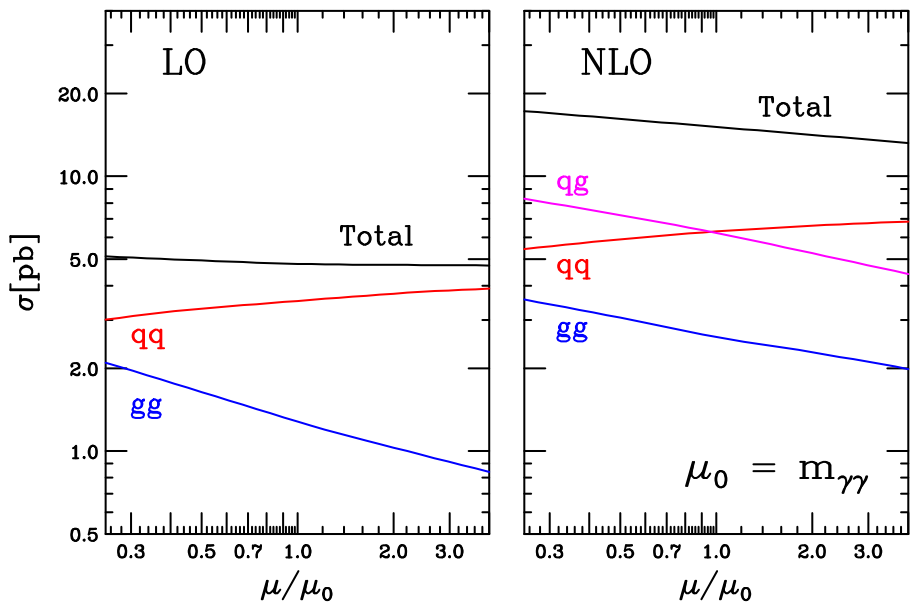}
\caption{Dependence of the LO and NLO diphoton cross sections at $\sqrt{s}=7$~TeV (in pb)
on the scale choice $\mu$.
We vary $\mu \equiv \mu_{R} = \mu_{F} = M_{F}$ about the central scale choice $\mu_0=m_{\gamma\gamma}$.
Total cross sections are shown in black whilst colours are used to denote the scale dependence of particular
initial states: quark-antiquark (red), quark-gluon (magenta), gluon-gluon (blue).
Photons are defined and isolated according to Eqs.~(\protect\ref{eq:gamgamstagcuts},~\protect\ref{eq:gamgamisol}).
\label{fig:gamgam_scale}
}
\end{center}
\end{figure}

For the cross sections presented so far we have chosen to set all
scales entering our calculation equal to the invariant mass of the two
photons, $\mu_R = \mu_F = M_F \equiv \mu_0$, with $\mu_0=m_{\gamma
\gamma}$.  In order to illustrate the impact of this choice, in
\fig~\ref{fig:gamgam_scale} we show the dependence of the theoretical
predictions on the common scale $\mu$ when it is varied by a factor
of four about $\mu_0$. In addition to the scale dependence of the total
predictions we also consider the variation with $\mu$ of the individual partonic
channels that appear at each order. Although the scale dependence of the
individual partonic processes is typically quite large (for example,
for the $q\overline{q}$ and $qg$ initiated processes at
NLO), the sum over all contributions is relatively scale-independent.
The LO cross section in particular has a tiny variation in this
range. The fact that the NLO corrections are large and not
reproducible for any choice of scale considered at LO serves as a
reminder that the scale variation is not indicative of the theoretical
uncertainty at that order.  We also note the large K-factor when going
from LO to NLO $(\sim 3.2)$, which is in stark contrast to the mild corrections
observed when imposing only basic acceptance cuts (c.f. \tab~\ref{gamgamtot}).
This difference can easily be understood from the nature
of the cuts in Eqs.~(\protect\ref{eq:gamgamstagcuts},~\protect\ref{eq:gamgamisol}).
For the Born and virtual contributions the staggered $p_T$ cut is effectively
a $p_T^{\gamma_2} > 40$~GeV cut due to the $2 \to 2$ kinematics.
Photons of $p_{T} < 40$~GeV can only be produced by fragmentation or real radiation
diagrams in which a parton is available to balance the staggered transverse momenta.
As a result these cuts strongly favour real radiation diagrams, a fact that is also
evident from the size of the $qg$ contribution in \fig~\ref{fig:gamgam_scale}.
Thus we would advocate using equal cuts on the photons, 
since they do not emphasize the role of higher order corrections. 

\begin{figure} 
\begin{center}
\includegraphics[width=12cm]{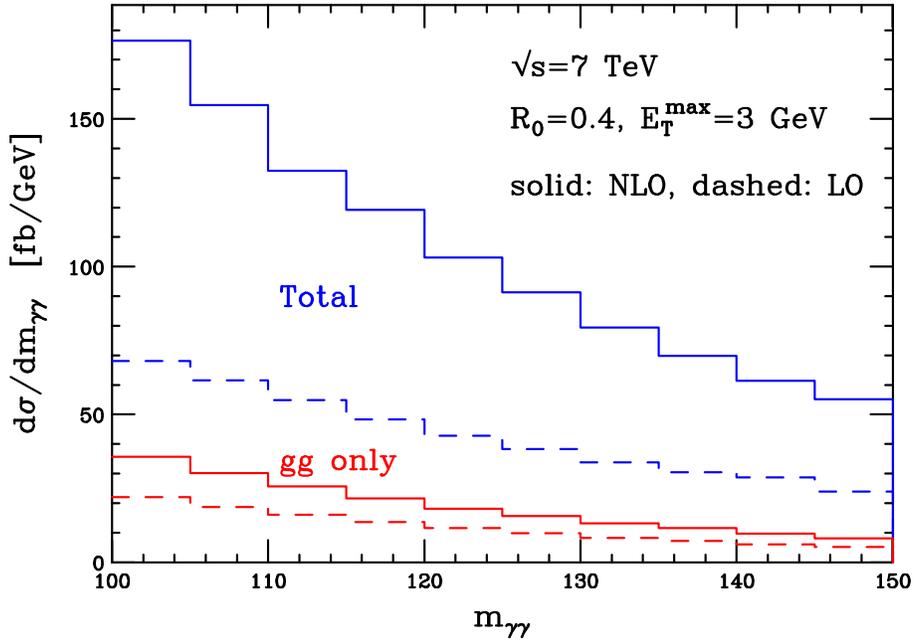}
\caption{The diphoton invariant mass distribution at $\sqrt{s}=7$~TeV (in fb/GeV).
We apply the staggered cuts described in the text and indicate LO results with dashed curves and
NLO results with solid curves. The two upper (blue) curves show the full predictions at a given order, while
the lower (red) curves indicate the gluon-gluon initiated contributions only.
\label{fig:mgamgam}}
\end{center}
\end{figure}

Finally, we consider predictions for the diphoton invariant mass distribution, a key ingredient in the
search for a light Higgs boson. Our results for $\sqrt{s}=7$~TeV and the cuts of
Eqs.~(\protect\ref{eq:gamgamstagcuts},~\protect\ref{eq:gamgamisol}) are shown in \fig~\ref{fig:mgamgam}.
It is clear that, in order to provide a good prediction for this distribution, one must
include not only the gluon-gluon initiated process but also the NLO corrections throughout.


\section{$W^\pm\gamma$ production}
\label{sec:Wgam}

\subsection{Description of the calculation}
The production of a $W$ boson and a photon proceeds at Born level via quark-antiquark annihilation,
\begin{equation}
q + \bar{q}^\prime \to W^\pm \gamma \;.
\end{equation}
This process was first calculated several decades ago~\cite{Brown:1979ux}, with the effect of
radiative corrections subsequently accounted for in ref.~\cite{Ohnemus:1992jn}. Since then the
subject has been revisited several times. A fully differential Monte Carlo implementation of the
NLO result is presented in ref.~\cite{DeFlorian:2000sg}, making use of the helicity amplitudes
calculated in ref.~\cite{Dixon:1998py}. Spin correlations in the decay of the $W$ boson are included
although no photon radiation from the lepton is allowed.
Electroweak corrections to this process~\cite{Accomando:2005ra} and NLO QCD corrections to the related
$W\gamma+$jet final state have also been computed~\cite{Campanario:2009um}.

\begin{figure}
\begin{center}
\includegraphics[scale=0.5,angle=270]{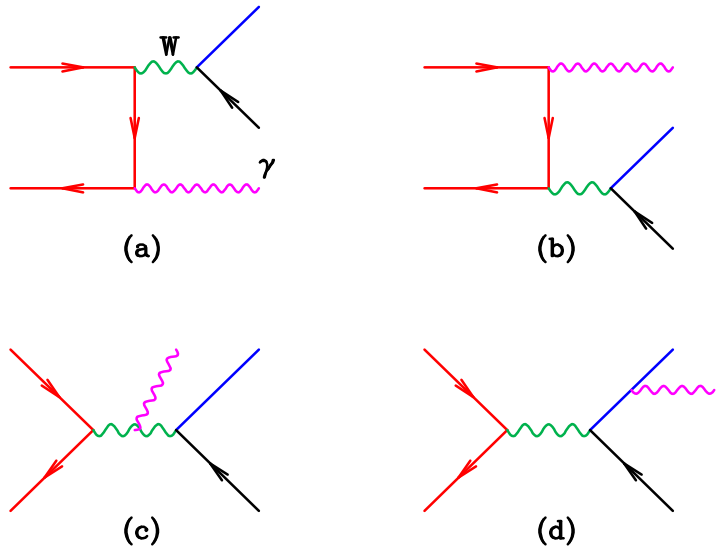}
\caption{Leading order diagrams for $W(\to \ell \nu)\gamma$ production. The diagrams (a),(b) and (c)
can be considered as radiation in the production process, while the final
diagram (d) corresponds to photon radiation from the lepton in the $W$ decay.
\label{Wgam}}
\end{center}
\end{figure}

In this section we present results using the current implementation of this process in MCFM.
The diagrams that contribute to this process at leading order are shown in \fig~\ref{Wgam}.
The next-to-leading order diagrams are obtained by dressing these diagrams with both virtual and real
gluon radiation. The contribution to the full amplitude arising from three of these diagrams is readily
obtained from the helicity amplitudes of ref.~\cite{Dixon:1998py}. The final diagram, including appropriate
dressings that are straightforward to compute, accounts for the
additional contribution from photon radiation in the leptonic decay of the $W$ boson. The resulting amplitude
retains full spin correlations in the decay.


\subsection{Results}

In order to define the final state for this process we apply a basic set of kinematic cuts,
\begin{equation}
p_{T}^\gamma > 10~\mbox{GeV} \;, \quad R_{\ell\gamma} > 0.7 \;,
\end{equation}
and demand that the photon be isolated as before, $R_0=0.4$ and $\Etmax=3$. In this subsection 
we consider $W$ bosons which decay leptonically. 
We do not apply any cuts to the leptons, except for the photon-lepton separation cut which ensures that the
photon-lepton collinear singularity is avoided. 
The resulting cross sections are given, as a function of $\sqrt{s}$, in \tab~\ref{Wgamcut}. 
\renewcommand{\baselinestretch}{1.6}
\begin{table}
 \begin{center}
 \begin{tabular}{|c|c|c|c|c|}
 \hline
 $\sqrt{s}$~[TeV] & $\sigma^{LO}(e^+\nu\gamma) $~[pb] &$\sigma^{NLO}(e^+\nu\gamma) $~[pb]&$\sigma^{LO}(e^-\overline{\nu}\gamma)$~[pb]&$\sigma^{NLO}(e^-\overline{\nu}\gamma) $~[pb]\\
 \hline
     7 &  23.02(6) &$  30.1(1)_{ -6\%}^{+ 5 \%} $&  15.46(5) &$  21.1(1)_{ -8\%}^{+ 4 \%}$ \\
     8 &  26.86(8) &$  35.1(2)_{ -7\%}^{+ 3 \%} $&  18.53(7) &$  24.6(1)_{ -6\%}^{+ 5 \%}$ \\
     9 &  30.62(8) &$  39.6(2)_{ -7 \%}^{+ 5 \%} $&  21.26(8) &$  28.4(2)_{ -6\%}^{+ 4 \%}$ \\
    10 &  34.6(1) &$  44.2(4)_{ -6 \%}^{+ 5 \%} $&  24.13(8) &$  32.2(2)_{ -8\%}^{+ 3 \%}$ \\
    11 &  38.4(1) &$  48.8(3)_{ -8 \%}^{+ 4 \%} $&  27.1(1) &$  35.7(2)_{ -6\%}^{+ 4 \%}$ \\
    12 &  42.2(1) &$  54.0(4)_{ -8 \%}^{+ 3 \%} $&  30.2(1) &$  39.4(2)_{ -6\%}^{+ 5 \%}$ \\
    13 &  45.9(1) &$  57.7(4)_{ -6 \%}^{+ 3 \%} $&  33.1(1) &$  43.6(3)_{ -8\%}^{+ 4 \%}$ \\
    14 &  49.8(1) &$  62.8(4)_{ -9 \%}^{+ 5 \%} $&  36.0(1) &$  47.4(3)_{ -8\%}^{+ 4 \%}$ \\
 \hline
 \end{tabular}
  \renewcommand{\baselinestretch}{1.0}
 \caption{Cross sections for $W(\to \ell \nu)\gamma$ production as a function of energy, using only the
cuts of Eq.~(\protect\ref{Wgambasic}). The cross sections are calculated including the effects of photon
radiation in the $W$ decay and the central values are obtained using $\mu_{R}=\mu_{F}=M_{F}=M_W$.
The uncertainty is derived from the scale dependence, as described in the text.}
 \label{Wgamcut}
 \end{center}
 \end{table}
\renewcommand{\baselinestretch}{1.0}
We present results for the LO and NLO cross sections for $e^+\nu\gamma$ and
$e^-\overline{\nu}\gamma$ separately. The cross sections have been
calculated using a central scale choice of $\mu_R=\mu_F=M_F=M_W$,
with upper and lower extrema obtained by evaluating the cross
section at $\{ \mu_R=M_W/2 , \mu_F=2M_W\}$ and $\{ \mu_R=2M_W ,
\mu_F=M_W/2\}$ respectively. The fragmentation scale is kept fixed at $M_W$
throughout since its variation does not lead to a significant change in our results
over the range of interest.
From this table we can readily extract our NLO prediction for the $W\gamma$ cross section
(summed over both $W^+$ and $W^-$) at current LHC operating energies with the cuts and isolation
described above, 
\begin{equation}
\sigma^{NLO}(pp\rightarrow W\gamma +X)\times
BR(W\rightarrow \ell\nu) = 51.2 \, ^{+2.3}_{-3.5}~\mbox{pb} \;.
\end{equation}
This is to be compared with a recently-reported cross section from the
CMS collaboration~\cite{EWK-10-008}. They find,
\begin{equation}
\sigma^{CMS}(pp\rightarrow W\gamma
+X)\times BR(W\rightarrow \ell\nu) = 55.9 \pm 5.0~\mbox{(stat)} \pm 5.0
~\mbox{(sys)} \pm 6.1~\mbox{(lumi)~pb} \;,
\end{equation}
in good agreement with the Standard Model expectation.

\begin{figure}
\begin{center}
\includegraphics[width=12cm]{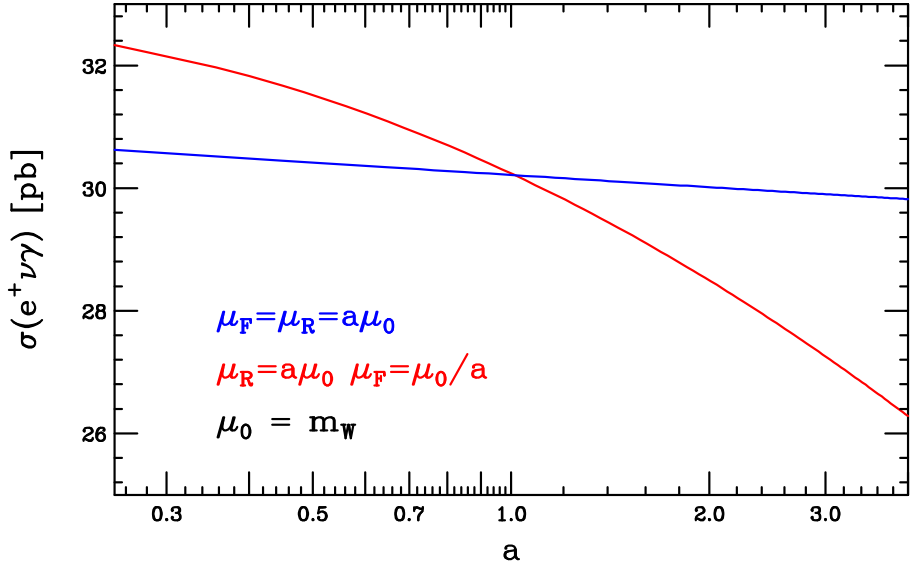}
\caption{Scale variation for $W^+(\to e^+ \nu)\gamma$ production, applying only the basic cuts of Eq.~(\protect\ref{Wgambasic}). 
For the red curve we vary the factorisation and renormalisation functions in opposite directions, whilst for the 
blue curve we vary them in the same direction. The fragmentation scale is kept fixed at $M_W$.} 
\label{fig:wscale}
\end{center}
\end{figure}

Varying the factorisation and renormalisation scales in
the manner that we have chosen requires some further justification since the normal
theoretical preference is to change them together in the same direction.
A comparison of these two choices, 
varying $\mu_F$ and $\mu_R$ in the same and opposite directions,
is shown in \fig~\ref{fig:wscale}.
We observe that there is essentially
no change in the NLO $e^+\nu\gamma$ cross section when the scales are varied in the same
direction, which is due to the qualitatively different behaviour of the contributing
partonic states. The $q\overline{q}$ initial state is dominated by
variations in the factorisation scale and grows with increasing $\mu_F$.
Conversely, the $gq$ initial state depends most strongly on the renormalisation scale
and decreases with increasing $\mu_R$. The combination of these two
initial states results in a very small net scale dependence. Since this is simply a
fortuitous cancellation and higher order corrections to the NLO cross section will
likely not be bracketed by this small scale variation, we choose to vary the scales in
opposite directions instead. We believe that this results in a more credible estimate of the
theoretical uncertainty of the calculation.

At LHC centre of mass energies the dominant contributions to the cross
sections that we have presented so far result from the radiation of a photon
from the lepton in the $W$ decay.  For studies of anomalous couplings of vector bosons to
photons, and for the observation of radiation zeros in rapidity
distributions, it is most useful to suppress this contribution. 
This is achieved by applying a cut on the transverse mass ($M_T$) of the
photon-lepton-MET system,  $M_{T} > 90$~GeV.  To investigate the role of lepton cuts on
the cross section and distributions we will first present results for the cross
section at $\sqrt{s}=7$~TeV including the $W$ decay for various sets of cuts.
The three sets of cuts that we will consider are,
\begin{eqnarray}
{\rm{Basic~Photon}} &:& p^{\gamma}_{T} > 10~{\rm{GeV}}, \; |\eta_{\gamma}| < 5, \; R_{\ell\gamma} > 0.7, \;R_0=0.4, \; E_{T}^{max}=3~\mbox{GeV}.
\label{Wgambasic} \\
M_T~{\rm{cut}} & : &  {\rm{Basic~Photon}} + M_T > 90~\mbox{GeV}.
\label{WgamMT} \\
{\rm{Lepton~cuts}} & : & M_T~{\rm{cut}} + E_T^{{\rm{miss}}} > 25~{\rm{GeV}}, \; p_{T}^{\ell} > 20~\mbox{GeV}, \; |\eta_{\ell}| < 2.5.
\label{Wgamfull}
\end{eqnarray}
For each of these sets of cuts we will perform our NLO calculation in two different ways.
In the first case  (no final-state radiation, ``No FSR'')  we will omit diagram \fig~\ref{Wgam}(d) corresponding to photon radiation
in the $W$ decay (and its appropriate NLO dressings). Such an approach is natural if one demands
that the lepton-neutrino system is produced exactly on the $W$ mass-shell. This corresponds to the
approach taken in ref.~\cite{DeFlorian:2000sg}. This constraint cannot be implemented physically.
For the second case (``Full'') we follow our usual procedure and include this diagram and
NLO counterparts.
The results are summarised in \tab~\ref{Wgamcuts}.
\renewcommand{\baselinestretch}{1.6}
\begin{table}
 \begin{center}
 \begin{tabular}{|c||c|c|c|c|c|}
 \hline
 Decay &Cuts & $\sigma^{LO}(e^+\nu\gamma) $ &$\sigma^{NLO}(e^+\nu\gamma) $&$\sigma^{LO}(e^-\overline{\nu}\gamma)$&$\sigma^{NLO}(e^-\overline{\nu}\gamma) $\\
 \hline
     No FSR &Basic $\gamma$ & 4.88 & 8.74  &  3.15 & 6.01   \\
  &  $M_{T}$ cut     		&  1.99	 	&3.78	&   1.26   	& 2.66    \\
  &    Lepton cuts 	         &  1.49			 &  	2.73 	&     0.86 &  1.77   	\\
  \hline
   Full &Basic $\gamma$ & 23.0 & 30.1		 &  	15.5		& 21.1  \\
  &  $M_{T}$ cut     & 2.12	 & 3.94	&   1.34 &  	 2.75      \\
  &    Lepton cuts     &  1.58 &  2.85	&  0.91        &   1.81	\\
     \hline
 \end{tabular}
 \renewcommand{\baselinestretch}{1.0}
 \caption{ 
  $W(\to \ell \nu)\gamma$ cross sections in picobarns at $\sqrt{s}=7$~TeV for the various scenarios described in detail in the text.
  Results in the upper half (``No FSR'') correspond to neglecting diagrams containing photon radiation
  in the $W$ decay, while the cross sections in the lower half (``Full'') include this effect.
  The cuts on the final state are specified in Eqs.~(\protect\ref{Wgambasic})--(\protect\ref{Wgamfull}).
  Statistical errors are $\pm 1$ in the final digit.
 \label{Wgamcuts}}
 \end{center}
 \end{table}
 \renewcommand{\baselinestretch}{1.0}

We observe that with just the basic cuts of Eq.~(\ref{Wgambasic}) the difference in predicted cross section
between the two calculations is very large. At NLO the full result is over three times larger than
the ``No FSR'' equivalent. As claimed earlier, applying the $M_T$ cut of Eq.~(\ref{WgamMT}) significantly
reduces this difference. The NLO cross section including radiation in the decay is about
$3$\% higher. The quantity that is most relevant experimentally corresponds to the full cuts given
in Eq.~(\ref{Wgamfull}). In that case the two calculations differ by at most $4$\% at NLO. 

\begin{figure}
\begin{center}
\includegraphics[width=12cm]{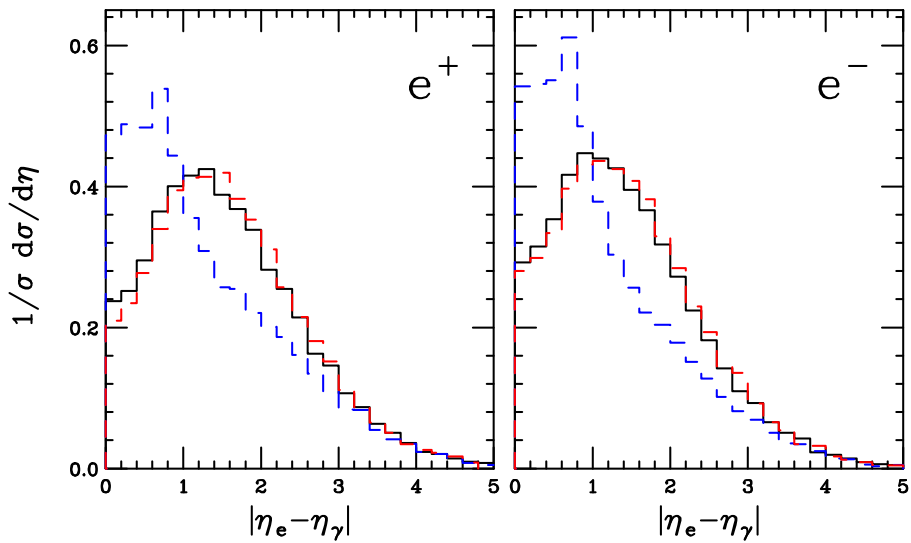}
\caption{NLO Predictions for the pseudorapidity difference between the charged lepton and the photon in $W(\to \ell \nu)\gamma$ 
events, for three different levels of the calculation. For all curves we apply the lepton cuts of Eq.~(\protect\ref{Wgamfull}).
The black curve represents the complete NLO prediction. 
The red dashed curve represents the NLO prediction in the case where no photon radiation is allowed
from the lepton (``No FSR''). The blue dashed curve has no cut on $M_T$, but keeps the cuts on the leptons.}
\label{fig:rapdif}
\end{center}
\end{figure}

Including the final state radiation of photons not only significantly increases
$W\gamma$ cross sections at the LHC, it also changes the character of the radiation zero~\cite{Mikaelian:1979nr}
that is present in the amplitude. The signature of the radiation amplitude zero can be seen
in the distribution of the pseudorapidity difference between the charged lepton and the photon.
Our predictions for this distribution, using $\mu_R=\mu_F=M_F=M_W$ and applying the
full lepton cuts of Eq.~(\ref{Wgamfull}), are shown in \fig~\ref{fig:rapdif}.
The dashed blue curve in \fig~\ref{fig:rapdif} represents the NLO rapidity difference with lepton cuts (Eq.~(\ref{Wgamfull})), but with no cut on
$M_T$ applied. We observe that the characteristic dip associated with the radiation zero has been completely filled in by the radiation of 
photons from the charged lepton. This is due to the fact that this configuration favours a collinear electron-photon pair so the rapidity difference between the 
two is usually small. Applying the $M_T$ cut (black curve) removes the majority of these configurations and the dip is restored. With the $M_T$ cut the NLO 
prediction from the full theory is similar to the result from the ``No FSR'' calculation (red curve).


\section{$Z\gamma$ production}
\label{sec:Zgam}

\subsection{Description of the calculation}
The production of a $Z$ boson and a photon primarily occurs through the Born process,
\begin{equation}
q + \bar{q} \to Z \gamma \;.
\end{equation}
The next-to-leading order corrections to this were computed in refs.~\cite{Ohnemus:1992jn,Baur:1997kz}
and later extended to the case of a decaying $Z$ boson in ref.~\cite{DeFlorian:2000sg}.
Electroweak corrections to this process have also been computed~\cite{Hollik:2004tm,Accomando:2005ra}.

A further contribution arises from the process,
\begin{equation}
g + g \to Z \gamma \;,
\end{equation}
which proceeds via a quark loop.
Since this contribution is finite it can be computed separately, as first detailed in
refs.~\cite{Ametller:1985di,vanderBij:1988fb}.
More recently this process has been computed including the leptonic decay of the $Z$
boson and other higher order contributions~\cite{Adamson:2002rm}.

The results presented in this section are obtained using our current implementation in MCFM
which is as follows. Strong corrections to the $q{\bar q}$ initiated process are fully
included, also allowing additional contributions from fragmentation processes.
The $gg$ process is included for five flavours of massless quarks. The contribution from
massive top quark loops is suppressed by $1/m_t^4$ and is therefore neglected.
We find agreement with the large top-mass limit of the results presented in ref.~\cite{vanderBij:1988fb},
where the full dependence on the top and bottom quark masses has been kept.

Analytic expressions for the $gg \to Z\gamma$ amplitudes that we include may be obtained from
existing results for $e^+ e^- \to 4$~partons~\cite{Bern:1997sc}, as described in
Appendix~\ref{app:amps}.

\subsection{Results}

We begin by assessing the impact of radiation in the decay of the $Z$ boson to charged leptons. 
As before we consider three sets of cuts to illustrate the difference between the two calculations.
These are:
\begin{eqnarray}
{\rm{Basic~Photon}} &:& m_{e^+e^-}>50~\mbox{GeV}, \; p^{\gamma}_{T} > 10~{\rm{GeV}}, \; |\eta_{\gamma}| < 5, \; R_{\ell\gamma} > 0.7, \nonumber
\\  && R_0=0.4, \; \Etmax=3~\mbox{GeV} \;. \label{Zgambasic} \\
M_{\ell\ell\gamma}~{\rm{cut}} & : &  {\rm{Basic~Photon}} + M_{\ell\ell\gamma}>100~\mbox{GeV} \;. \label{ZgamMT} \\
{\rm{Lepton~cuts}} & : & M_{\ell\ell\gamma}~{\rm{cut}} + p_{T}^{\ell} > 20~\mbox{GeV}, \; |\eta_{\ell}| < 2.5 \;. \label{Zgamfull} 
\end{eqnarray}
The first set of cuts, Eq.~(\ref{Zgambasic}), is very similar to the basic cuts for the $W\gamma$ process (Eq.~(\ref{Wgambasic})) but with an additional
dilepton invariant mass cut in order to select real $Z$ events. The cut on the transverse mass $M_T$ has been replaced with a cut on the invariant mass 
of the photon+leptons system. This reflects the fact that for $Z(\rightarrow\ell^+\ell^-)\gamma$ production all of the final state is reconstructed. Also the value of this mass cut must be slightly higher than the equivalent for
$W\gamma$ (Eq.~(\ref{WgamMT})), to reflect the $\sim 10$~GeV higher mass of the $Z$ boson. The lepton cuts are identical, without of course
any requirement on the missing transverse energy. These cuts are motivated by an early CMS study~\cite{EWK-10-008}.
Our results for $\sqrt{s}=7$~TeV, shown in \tab~\ref{Zgamcuts}, indicate that the $M_{\ell\ell\gamma}$ cut is reasonably effective at removing the contribution
to the cross section from photons in the $Z$ decay. In the presence of the full lepton cuts,
given in Eq.~(\ref{Zgamfull}), including photon radiation in the decay increases the cross section by about $15$\% at
NLO. This is a larger difference than for $W\gamma$ production, which is to be expected since radiation may occur from both decay
products.
\renewcommand{\baselinestretch}{1.6}
\begin{table}
 \begin{center}
 \begin{tabular}{|c||c|c|c|}
 \hline
 Decay &Cuts & $\sigma^{LO}(e^+e^-\gamma) $ &$\sigma^{NLO}(e^+e^-\gamma)$\\
 \hline
   No FSR
  &Basic $\gamma$ 		& 1.67(0) & 2.33(0)  \\
  &  $M_{\ell\ell\gamma}$ cut     		& 	1.67(0) 	& 2.29(0)		 \\
  &    Lepton cuts 	        		&  0.82(0)	 	& 1.17(0) 		\\
  \hline
 Full 
  &Basic $\gamma$ 		& 7.84 	& 	9.83  \\
  &  $M_{\ell\ell\gamma}$ cut     		& 	2.08(0)	 & 	2.81	\\
  &   Lepton cuts     		& 	0.99(0)	& 	1.39(0) 	\\
     \hline
 \end{tabular}
 \renewcommand{\baselinestretch}{1.0}
 \caption{
  $Z(\to e^+ e^-)\gamma$ cross sections in picobarns at $\sqrt{s}=7$~TeV for the various scenarios described in detail in the text.
  Results in the upper half (``No FSR'') correspond to neglecting diagrams containing photon radiation
  in the $Z$ decay, while the cross sections in the lower half (``FSR'') include this effect.
  The cuts on the final state are specified in Eqs.~(\protect\ref{Zgambasic})--(\protect\ref{Zgamfull}).
  Statistical errors, unless otherwise indicated, are $\pm 1$ in the final digit.
 \label{Zgamcuts}}
 \end{center}
 \end{table}
 \renewcommand{\baselinestretch}{1.0}

We now turn to the issue of the dependence of the cross section on the centre-of-mass energy $\sqrt{s}$ and the estimation
of the theoretical uncertainty from scale variation. As is the case for
the $W\gamma$ cross sections of the previous section, we find that varying all scales by a factor of two about the central
value of $M_Z$ results in a very small scale dependence. This is a result of the same accidental cancellation between the
scaling behaviours of the component partonic cross sections. The scale uncertainties are therefore obtained by keeping $M_F=M_Z$
(since the fragmentation contribution is itself very small) and using $\{\mu_R = M_Z/2, \mu_F=2 M_Z\}$ and
$\{\mu_R = 2M_Z, \mu_F=M_Z/2\}$ for the upper and lower extrema respectively. Our results are shown in \tab~\ref{Zgammatot}. 
\renewcommand{\baselinestretch}{1.6}
 \begin{table}
 \begin{center}
 \begin{tabular}{|c|c|c|}
 \hline
 $\sqrt{s}$~[TeV] & $\sigma^{LO}(e^+e^-\gamma) $~[pb] &$\sigma^{NLO}(e^+e^-\gamma) $~[pb] \\
 \hline
     7 &   7.84(1) &$   9.83(1)_{ -4.7\%}^{+   3.6\%}  $\\
     8 &   9.23(1) &$  11.48(1)_{ -5.1\%}^{+   3.5\%}  $\\
     9 &  10.65(2) &$  13.10(1)_{ -5.4\%}^{+   3.6\%}  $\\
    10 &  12.10(2) &$  14.72(1)_{ -5.7\%}^{+   3.7\%} $ \\
    11 &  13.56(2) &$  16.38(2)_{ -6.1\%}^{+   3.6\%} $ \\
    12 &  15.01(3) &$  18.00(2)_{ -6.2\%}^{+   3.5\%} $ \\
    13 &  16.50(3) &$  19.61(2)_{ -6.6\%}^{+   3.6\%} $ \\
    14 &  17.97(3) &$  21.20(2)_{ -6.6\%}^{+   3.7\%} $ \\
 \hline
 \end{tabular}
 \renewcommand{\baselinestretch}{1.0}
\caption{
Cross sections for $Z(\to e^+ e^-)\gamma$ production as a function of energy, using only the
cuts of Eq.~(\protect\ref{Zgambasic}). The cross sections are calculated including the effects of photon
radiation in the $Z$ decay and the central values are obtained using $\mu_{R}=\mu_{F}=M_{F}=M_Z$.
The uncertainty is derived from the scale dependence, as described in the text.
\label{Zgammatot}
}
 \end{center}
 \end{table}
 \renewcommand{\baselinestretch}{1.0} 
Finally, we can once again compare our NLO prediction for the total $Z\gamma$ cross section to a measurement already
made at the LHC. From \tab~\ref{Zgammatot} we see that our NLO prediction for the cross section at $7$~TeV and using
the cuts of Eq.~(\ref{Zgambasic}) is,
\begin{equation}
\sigma^{NLO}(Z\gamma)\times BR(Z\rightarrow \ell^-\ell^+) = 9.83 \, _{-0.46}^{+0.35}~\mbox{pb} \;.
\end{equation}
The corresponding result reported by the CMS collaboration is~\cite{EWK-10-008},
\begin{equation}
\sigma^{CMS}(Z\gamma)\times BR(Z\rightarrow \ell^-\ell^+) = 9.3 \pm 1.0~\mbox{(stat)} \pm 0.6~\mbox{(syst)} \pm 1.0~\mbox{(lumi)~pb}
\end{equation}
which is already in good agreement within errors. 

\begin{figure}
\begin{center}
\includegraphics[width=10cm]{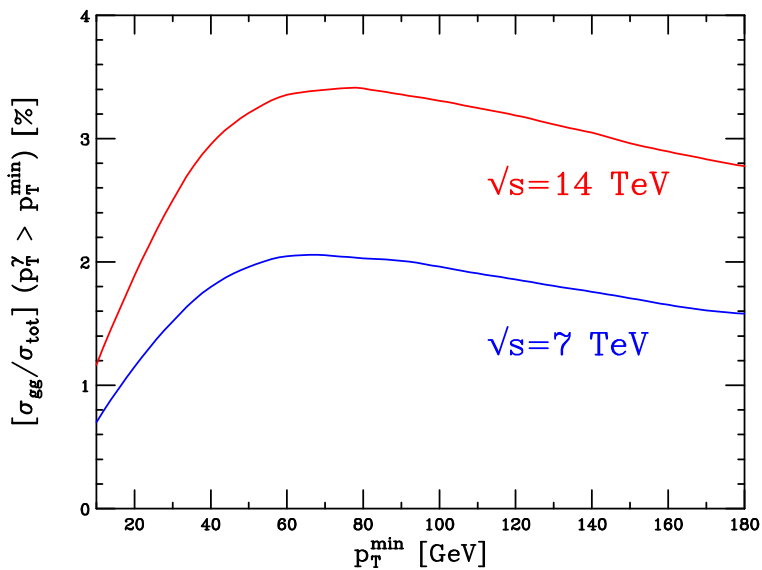}
\caption{The percentage of the $Z(\to e^+ e^-)\gamma$ cross section using the cuts of Eq.~(\protect\ref{Zgamfull}) 
contributed by the gluon-gluon initiated
diagrams, as a function of the minimum photon $p_T$ allowed in the events. The upper (red)
curve is for $\sqrt{s}=14$~TeV while the lower (blue) curve corresponds to  $\sqrt{s}=7$~TeV.
\label{Zgam_gg}}
\end{center}
\end{figure}

We conclude with an investigation of the importance of the gluon-gluon contribution in phenomenological
studies. We shall use the full set of cuts given in Eq.~(\ref{Zgamfull}) as indicative of
the appropriate experimental acceptance at the LHC. In that case we note that the relative effect
of adding these diagrams is small since the cross section is dominated by regions of low
$p^\gamma_T$ that are enhanced for the $q\bar q$ process but not for the loop-induced $gg$
diagrams. However, as one moves to moderate values of $p^\gamma_T$ one would expect
the relative size of the gluon-gluon contribution to grow. This is exactly the behaviour that
we observe in \fig~\ref{Zgam_gg}, with the $gg$ fraction falling again at higher $p^\gamma_T$ due to
the behaviour of the parton fluxes.
We also see that, as expected, the gluon-gluon contribution is more important at 14~TeV,
although it is still at most 3.5\% of the total NLO cross section.

\section{$WW$ production}
\label{sec:WW}

\subsection{Description of the calculation}
The production of a pair of $W$ bosons is an important channel, in part because of its role
as a background to Higgs boson searches in which the Higgs decays into $W$ pairs.
The total cross section for the process,
\begin{equation}
q + \bar{q} \to W^+ W^- \;,
\end{equation}
was first calculated in the Born approximation in ref.~\cite{Brown:1978mq}, with strong corrections to it
given in refs.~\cite{Ohnemus:1991kk,Frixione:1993yp,Ohnemus:1994ff}. 
These processes are included in MCFM at NLO
using the one-loop amplitudes presented in ref.~\cite{Dixon:1998py}. Phenomenological NLO results for the Tevatron
and the LHC operating at $\sqrt{s}=14$~TeV have been presented in refs.~\cite{Campbell:1999ah,Dixon:1999di}. NLO results are also available
for the processes $W^+W^-+$~jet~\cite{Campbell:2007ev,Dittmaier:2009un}, $W^+W^++2$~jets~\cite{Melia:2010bm} 
and $W^+W^-+2$~jets~\cite{Melia:2011dw}.

The contribution for the process,
\begin{eqnarray}
g+g\rightarrow W^+W^- \;,
\end{eqnarray}
was first calculated in refs.~\cite{Dicus:1987dj,Glover:1988fe}. 
A more recent analysis of these contributions is given in ref.~\cite{Binoth:2005ua}
where off-shell effects of the vector bosons and their subsequent decays are taken into
account. Finally, the most complete analysis of these contributions to date is given in
ref.~\cite{Binoth:2006mf} where the effect of massive quarks circulating 
in the loop is included. The authors find that the effect of including the third $(t,b)$
isodoublet increases the gluon-gluon contribution by at most a factor of $12\%$ at the $14$~TeV LHC.

The results presented in this section are obtained using our current implementation in MCFM
which is as follows. Strong corrections to the $q{\bar q}$ initiated process are fully
included, with additional contributions from singly resonant diagrams as described in
ref.~\cite{Campbell:1999ah}. Since the contribution from the $(t,b)$ isodoublet to the $gg$
initiated process is small -- certainly much smaller than the residual uncertainty resulting
from the $O(\alpha_s^2)$ nature of the contribution -- the $gg$ process is included for two
massless generations only. Our results for the $gg$ process are in complete agreement with the equivalent two generation
results presented in ref.~\cite{Binoth:2006mf}. As can be seen from Table 2 therein, the final cross
section summed over $q\bar q$ and $gg$ channels is smaller than the three generation result by $0.5$\%.

The inclusion of the $gg$ contribution with massless quarks in the loop is straightforward.
The amplitudes can be obtained by simply recycling compact analytic expressions for certain contributions to the
process $e^+ e^- \to \mbox{4 partons}$ presented in ref.~\cite{Bern:1997sc}. The precise relations
are given in Appendix~\ref{app:amps}.

\subsection{Results}

We begin our discussion of $WW$ production by presenting the cross section as a function of $\sqrt{s}$ in \tab~\ref{WWtot}.
\renewcommand{\baselinestretch}{1.6}
 \begin{table}
 \begin{center}
 \begin{tabular}{|c|c|c|}
 \hline
 $\sqrt{s}$~[TeV] & $\sigma^{LO}(W^+W^-) $~[pb] &$\sigma^{NLO}(W^+W^-) $~[pb] \\
 \hline
     7 &  29.51(1) &$  47.04(2)_{ -3.2\%}^{+  4.3\%} $ \\
     8 &  35.56(1) &$  57.25(2)_{ -2.8\%}^{+  4.1\%} $ \\
     9 &  41.75(2) &$  67.82(3)_{ -2.8\%}^{+  3.8\%} $ \\
    10 &  48.07(2) &$  78.70(3)_{ -2.5\%}^{+  3.6\%} $ \\
    11 &  54.53(2) &$  89.80(4)_{ -2.5\%}^{+  3.3\%} $ \\
    12 &  61.10(3) &$ 101.14(5)_{ -2.4\%}^{+  3.1\%} $ \\
    13 &  67.74(3) &$ 112.64(5)_{ -2.3\%}^{+  3.0\%} $ \\
    14 &  74.48(4) &$ 124.31(6)_{ -2.0\%}^{+  2.8\%} $ \\
 \hline
 \end{tabular}
 \renewcommand{\baselinestretch}{1.0}
 \caption{ Total cross sections for $WW$ production as a function of energy. Renormalisation and factorisation scales are set to $M_W$. Upper and lower 
limits are obtained by varying the scales by a factor of two in each direction. Vector bosons are kept on-shell, with no branching ratios applied }
 \label{WWtot}
 \end{center}
 \end{table}
\renewcommand{\baselinestretch}{1.0}
The values are obtained by evaluating the cross  section with a central scale choice of $\mu_R=\mu_F=M_W$. Scale dependence is
illustrated by presenting percentage deviations from the central value as the scales are changed simultaneously by a factor of
two in each direction. The $W$ bosons are kept exactly on-shell and no decays are included for the cross sections presented in this
table. We note that as for the other diboson cross sections the NLO corrections
are typically large, enhancing the LO prediction by about a factor of $1.6$. 
From the table, our NLO prediction for the total $WW$ cross section at $\sqrt{s}=7$~TeV is,
\begin{eqnarray}
\sigma^{NLO}=47.0\,^{+2.0}_{-1.5} \,\,\rm{pb} \;. 
\end{eqnarray}
Although the general-purpose detectors at the LHC have collected only a handful of such events, both ATLAS~\cite{Collaboration:2011kk}
and CMS~\cite{Chatrchyan:2011tz} have already reported first measurements of this cross section. They find, 
\begin{eqnarray}
\sigma^{ATLAS}(WW)&=&41^{+20}_{-16}\,\rm{ (stat)} \pm 5 \,\, {\rm{ (syst)}}\,\,  \pm 1 \,\, {\rm{(lumi) \,\, pb}} \;, \\ 
\sigma^{CMS}(WW)&=& 41.1\pm 15.3 \,\rm{(stat)}\, \pm 5.8 \,{\rm{(syst)}} \pm 4.5 \,\, {\rm{(lumi) \,\, pb}} \;,
\end{eqnarray}
both of which are clearly compatible with the SM prediction.

A measurement of the $WW$ cross section at the LHC typically involves a jet veto to reduce the abundant top
background~\cite{Chatrchyan:2011tz,Collaboration:2011kk}.
Since a jet-veto can change the relative size of the NLO corrections we will study the dependence of the NLO cross section
on the transverse momentum scale used to veto jets, $p_{T}^{veto}$. For our purposes here we define the jet veto as a veto on all
jets with $p_{T} >p_{T}^{veto}$ that satisfy the rapidity requirement $|\eta_{j}| < 5$.
It is useful to consider the action of the jet veto under two sets of cuts, 
\begin{eqnarray}
{\rm{Basic}} \,\,\,WW &:& \,\, p^{\ell}_{T} > 20~{\rm{GeV}}, \; |\eta^{\ell}|<2.5, \; E_{T}^{miss}> 20~{\rm{GeV}} \;,
\label{eq:WWbas} \\
{\rm{Higgs}} \,\, &:& {\rm{Basic}} \,\,\,WW  + m_{\ell\ell} < 50~{\rm{GeV}}, \; \Delta \phi_{\ell\ell} < 60^{\circ}, \nn \\
 && \quad p^{\ell,max}_{T} > 30~{\rm{GeV}} , \; p^{\ell,min}_{T} > 25~{\rm{GeV}} \;.
\label{eq:WWfull}
\end{eqnarray}
These cuts are typical of those used at the LHC to measure the total
$WW$ cross section, (with the additional application of a jet-veto)
and those used to search for a Higgs boson~\cite{Chatrchyan:2011tz}. The precise nature of the
Higgs search cuts are dependent on the putative mass of the Higgs boson so
here we have selected a set used for $m_H = 160$ GeV, when the decay
to $WW$ is largest.

\begin{figure} 
\begin{center}
  \renewcommand{\baselinestretch}{1.0}
\includegraphics[width=12cm]{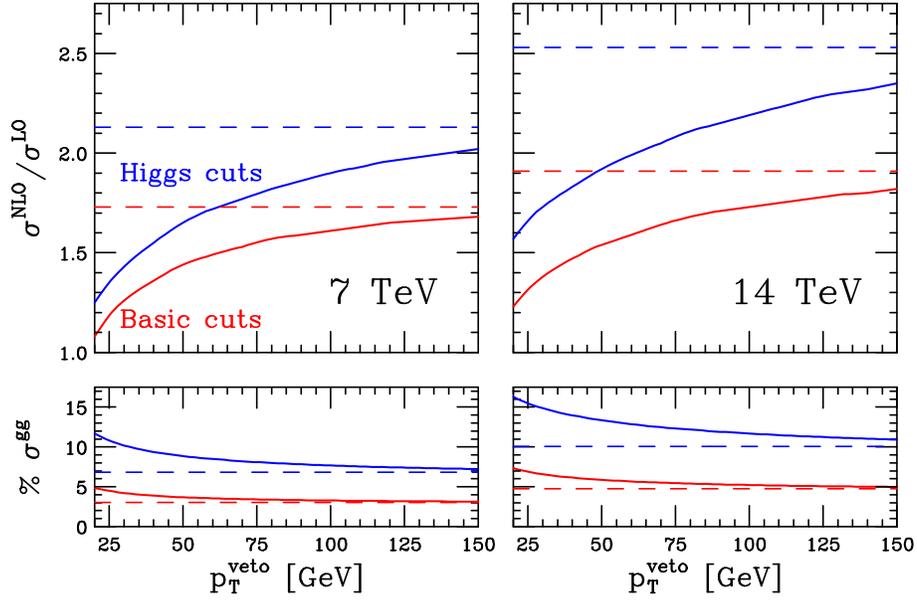}
\caption{The ratio of NLO to LO (upper) and the percentage of the NLO cross section from the $gg$ initial state (lower)
for $WW\rightarrow e^+\mu^-\nu_e\overline{\nu}_{\mu}$ production, as a function of the jet-veto $p^{veto}_T$.
Results are shown using the basic cuts of Eq.~(\protect\ref{eq:WWbas}) (upper, blue curves) and the Higgs search cuts of
Eq.~(\protect\ref{eq:WWfull}) (lower, red curves).
The NLO to LO ratio and gluon percentage with no veto 
applied are shown as dashed lines on the plot. The dashed lines are thus the asymptotic values of the solid curves.
\label{fig:WWjetveto}}
\end{center}
\end{figure}

The ratio of the NLO to LO cross sections, as a
function of $p_T^{veto}$ and for the two sets of cuts above, is shown
in the upper panels of \fig~\ref{fig:WWjetveto}.  Since the $gg$
initiated contribution does not contain any final state partons it is
unaffected by the jet-veto at this order.  As a result the
relative importance of this contribution increases when a jet-veto is
applied. We illustrate this by presenting the ratio
$\sigma({gg})/\sigma^{NLO}$ in the lower panels of
\fig~\ref{fig:WWjetveto}.
As expected, the application of a jet-veto can reduce the $K$-factor
considerably. For instance, applying a jet veto at $p_T^{veto}=20$~GeV
reduces the inclusive $K$-factor by around 40\%. From
\fig~\ref{fig:WWjetveto} we also observe that the Higgs cuts increase
both the impact of NLO corrections and the gluon initiated
contributions. The importance of the gluon initiated terms for Higgs searches has
been observed in previous studies~\cite{Binoth:2008pr}. Indeed these
studies have shown that, at $\sqrt{s}=14$ TeV and with stricter cuts
than those of Eq.~(\ref{eq:WWfull}), the $gg$ contributions can be as
large as 30\% of the NLO cross section~\cite{Binoth:2008pr}. At
$\sqrt{s}=7$ TeV and with cuts appropriate for this center of mass
energy we find that the $gg$ contribution is around 12\% of the total
NLO cross section with a jet veto of $20$~GeV, as shown in the lower
panel of \fig~\ref{fig:WWjetveto}.  The values of the asymptotic
limits of the jet veto curves shown in Fig~\ref{fig:WWjetveto},
corresponding to the $K$-factor and $gg$ percentage with no veto
applied, are collected in \tab~\ref{WWnums}. For completeness we also
include the corresponding predictions for the cross sections at LO and
NLO.
\renewcommand{\baselinestretch}{1.6}
\begin{table}
 \begin{center}
 \begin{tabular}{|c|c|c|c|c|}
 \hline
 $\sqrt{s}$~[TeV] and cuts & $\sigma^{LO}(e^+\mu^-\nu_e\overline{\nu}_{\mu}) $ [fb]
 &$\sigma^{NLO}(e^+\mu^-\nu_e\overline{\nu}_{\mu}) $ [fb] & $K$-factor & {\%} gg \\
 \hline
     7 	(Basic)			&   	144 	 &  249	         &  1.73	& 	3.05	\\ 
     7 	(Higgs)			&   	7.14  	 & 15.19         &  2.13	& 	6.85	\\ 
\hline     
    14 (Basic)			&        296 	 &   566 	 &  1.91	&  	4.73	\\
     14	(Higgs)			&   	13.7 	 &  34.7         &  2.53	& 	10.09	\\ 
   \hline
 \end{tabular}
 \renewcommand{\baselinestretch}{1.0}
 \caption{$W^+(\rightarrow e^+\nu_e)W^-(\rightarrow\mu^-\overline{\nu}_{\mu})$ cross sections in femtobarns at LO and NLO, the
 resulting $K$-factor 
and the percentage of the NLO cross section originating from gluon initiated contributions. Results are shown for the Basic
 (Eq.~(\protect\ref{eq:WWbas})) and Higgs (Eq.~(\protect\ref{eq:WWfull})) cuts.}
 \label{WWnums}
 \end{center}
 \end{table}
\renewcommand{\baselinestretch}{1.0}


\section{$W^\pm Z$ production}
\label{sec:WZ}

\subsection{Description of the calculation}
The production of a $WZ$ pair proceeds at LO through the process,
\begin{equation}
q + \bar{q}^\prime \to W^\pm Z \;.
\end{equation}
This process was first calculated to NLO in refs.~\cite{Ohnemus:1991gb,Frixione:1992pj}.
The inclusion of subsequent $W$ and $Z$ decays was added in
ref.~\cite{Ohnemus:1994ff}, partially including the effect of spin
correlations. The full effect of spin correlations at NLO was later
examined in refs.~\cite{Campbell:1999ah,Dixon:1999di}, using the virtual
amplitudes of ref.~\cite{Dixon:1998py}. The QCD corrections to the process in
which an additional jet is radiated are also now known~\cite{Campanario:2010hp}.

The results presented in this section are obtained using the same implementation
in MCFM as described in ref.~\cite{Campbell:1999ah}. In particular we include
contributions from singly resonant diagrams that can be significant when one
of the bosons is off-shell. The program includes both the contribution of a $Z$ 
and a virtual photon, when considering the decay to charged leptons.
We note that charge conservation precludes any contribution
from gluon-gluon diagrams of the type previously discussed for $WW$
production.

\subsection{Results}
The production of $WZ$ pairs provides a valuable test of the triple gauge boson
couplings (for a recent example, see for instance ref.~\cite{Abazov:2010qn})
and is a source of SM background events, for example in SUSY trilepton
searches~\cite{2009arXiv0901.0512T,Ball:2007zza}.
It is also a background for SM Higgs searches in the case of leptonic decays,
when one of the leptons is missed.
In order to normalize the $WZ$ background to such searches, in
\tab~\ref{WZtot} we show results for the total cross section for 
$WZ$ production at the LHC, as a function of the centre of mass energy.
\renewcommand{\baselinestretch}{1.6}
\begin{table}
 \begin{center}
 \begin{tabular}{|c|c|c|c|c|}
 \hline
 $\sqrt{s}$~[TeV] & $\sigma^{LO}(W^+Z) $~[pb] &$\sigma^{NLO}(W^+Z) $~[pb]&$\sigma^{LO}(W^-Z)$~[pb]&$\sigma^{NLO}(W^-Z) $~[pb]\\
 \hline
     7 &   6.93(0) &$  11.88(1)_{ -4.2\%}^{+ 5.5\%} $&    3.77(0) &$   6.69(0)_{ -4.3\%}^{+ 5.6 \%}$ \\
     8 &   8.29(1) &$  14.48(1)_{ -4.0\%}^{+ 5.2\%} $&    4.65(0) &$   8.40(0)_{ -4.1\%}^{+ 5.4 \%}$ \\
     9 &   9.69(1) &$  17.18(1)_{ -3.9\%}^{+ 4.9\%} $&    5.57(0) &$  10.21(0)_{ -3.9\%}^{+ 5.0 \%}$ \\
    10 &  11.13(1) &$  19.93(1)_{ -3.7\%}^{+ 4.8\%} $&    6.53(0) &$  12.11(1)_{ -3.7\%}^{+ 4.8 \%}$ \\
    11 &  12.56(1) &$  22.75(2)_{ -3.5\%}^{+ 4.5\%} $&    7.51(0) &$  14.07(1)_{ -3.6\%}^{+ 4.6 \%}$ \\
    12 &  14.02(1) &$  25.63(2)_{ -3.3\%}^{+ 4.3\%} $&    8.51(1) &$  16.10(1)_{ -3.4\%}^{+ 4.4 \%}$ \\
    13 &  15.51(2) &$  28.55(2)_{ -3.2\%}^{+ 4.1\%} $&    9.53(1) &$  18.19(1)_{ -3.3\%}^{+ 4.1 \%}$ \\
    14 &  16.98(2) &$  31.50(3)_{ -3.0\%}^{+ 3.9\%} $&   10.57(1) &$  20.32(1)_{ -3.1\%}^{+ 3.9 \%}$ \\
 \hline
 \end{tabular}
  \renewcommand{\baselinestretch}{1.0}
 \caption{
 Total cross sections for $WZ$ production as a function of energy. Renormalisation and
 factorisation scales are set equal to the average mass  of the $W$ and $Z$ i.e.
 $\mu_R=\mu_F=(M_W+M_Z)/2$. Upper and lower percentage deviations are obtained by varying the
 scales around the central scale by a factor of two. The vector boson are kept
 on-shell, with no decays included.}
 \label{WZtot}
 \end{center}
 \end{table}
Both renormalisation and factorisation scales are set to the mean vector
boson mass, $(M_W+M_Z)/2$. Since the LHC is a proton-proton machine, the $W^+Z$
and $W^-Z$ cross sections are not equal, with the ratio $\sigma^{NLO}(W^-Z)/\sigma^{NLO}(W^+Z)$ varying
between $0.56$ (for $\sqrt{s}=7$~TeV) and $0.65$ (at $14$~TeV).
\renewcommand{\baselinestretch}{1.0}

The study of boosted objects at the LHC has potential as an additional handle
on searches for new physics, such as a Higgs boson~\cite{Butterworth:2008iy} or
supersymmetric particles~\cite{Abdesselam:2010pt}. A possible first step for such searches
would be to validate the method by performing a similar analysis for 
known Standard Model particles. In this regard, $WZ$ production would be a natural
proxy for associated Higgs production, $WH$, where the decay of the $Z$ boson to
a bottom quark pair is a stand-in for the decay of a light Higgs boson.

\begin{figure} 
\begin{center}
\includegraphics[width=12cm]{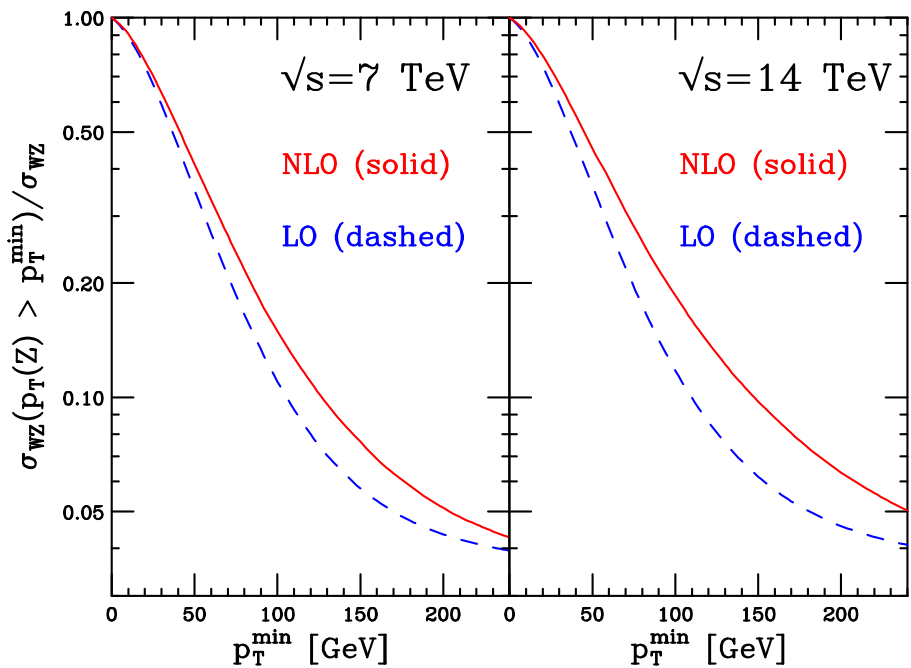}
\caption{The fraction of the total $WZ$ cross section surviving a cut on the $Z$ boson
transverse momentum, $p_T(Z) > p_T^{\rm min}$, at $\sqrt{s}=7$~TeV (left panel) and
$\sqrt{s}=14$~TeV (right panel). The NLO prediction is shown as a solid red curve and the LO
one is dashed blue.
\label{fig:WZpt}
}
\end{center}
\end{figure}

To this end, in \fig~\ref{fig:WZpt} we show the cross section for $WZ$ production
as a function of the minimum $Z$ boson transverse momentum, at $\sqrt{s}=7$ and $\sqrt{s}=14$~TeV. 
We observe that the number of boosted $Z$ bosons is sensitive both to the order in perturbation theory
and  the operating energy of the LHC. At both centre of mass energies the NLO prediction increases the
number  of high-$p_{T}$ $Z$ bosons, although the effect is larger at $14$ TeV. To emulate a typical
boosted Higgs search, with a $p_T$ cut at $200$~GeV, one thus retains about $5\%$ of the
total NLO cross section -- similar to the fraction for a putative Higgs signal.


\section{$ZZ$ production}
\label{sec:ZZ}

\subsection{Description of the calculation}
Although the production of $Z$ pairs is much smaller than the other diboson
cross sections considered above, it still plays an important role as principal
background to searches for a Higgs boson around the $Z$ pair threshold.
The NLO corrections to the process,
\begin{equation}
q + \bar{q} \to Z Z \;,
\end{equation}
were first calculated in refs.~\cite{Ohnemus:1990za,Mele:1990bq}, while the
inclusion of spin correlations in the decays and phenomenology for the
Tevatron and $14$~TeV LHC was presented in refs.~\cite{Campbell:1999ah,Dixon:1999di}.
Contributions from a gluon-gluon initial state,
\begin{equation}
g + g \to Z Z \;,
\end{equation}
were first considered in refs.~\cite{Dicus:1987dj,Glover:1988rg}.
The inclusion of leptonic decays of the Z bosons was examined in
refs.~\cite{Matsuura:1991pj,Zecher:1994kb} and later investigated in the context of
Higgs boson searches~\cite{:2008uu,Binoth:2008pr}. Furthermore, NLO results are
also available for the closely-related $ZZ$+jet process~\cite{Binoth:2009wk}.

The results presented in this section are obtained using our current implementation in MCFM
which is as follows. Strong corrections to the $q{\bar q}$ initiated process include singly-resonant
contributions -- a slight extension of the results presented in ref.~\cite{Campbell:1999ah} --
and the $gg$ process is included for five massless flavours. The contribution from
massive top quark loops is suppressed by $1/m_t^4$ and is therefore neglected. This
approximation results in gluon-gluon contributions that are $1\%$ lower than those reported
in refs.~\cite{:2008uu,Binoth:2008pr}, where the effects of massive top and bottom loops
are included.\footnote{We note that, when restricting our calculation to four massless flavours,
our results are in complete agreement with the equivalent cross section quoted in
ref.~\cite{:2008uu}.}
Finally, we observe that all our amplitudes also contain contributions from virtual photons.

The basic amplitudes entering the calculation of the $gg$ contribution are simply related
to those already discussed for the $gg \to WW$ process and are detailed in Appendix~\ref{app:amps}.

\subsection{Results}

We first present results for the dependence of the total cross section for $ZZ$ production as a
function of $\sqrt{s}$. As was the case for similar studies in previous sections
we keep the $Z$ bosons on-shell and do not include any decays. We choose a central scale choice
of $\mu_R=\mu_F=M_Z$ and vary this central scale by a factor of two in each direction to obtain
an estimate of the theoretical uncertainty. Our results are shown in \tab~\ref{ZZtot}.
\renewcommand{\baselinestretch}{1.6}
  \begin{table}
 \begin{center}
 \begin{tabular}{|c|c|c|}
 \hline
 $\sqrt{s}$~[TeV] & $\sigma^{LO}(ZZ) $~[pb] &$\sigma^{NLO}(ZZ) $~[pb] \\
 \hline
     7 &   4.17(0) &$   6.46(0)_{  -3.3\%}^{+  4.7\%} $ \\
     8 &   5.06(0) &$   7.92(0)_{  -3.0\%}^{+  4.7\%}  $\\
     9 &   5.98(0) &$   9.46(0)_{  -3.0\%}^{+  4.3\%}  $\\
    10 &   6.93(0) &$  11.03(0)_{  -2.9\%}^{+  4.1\%} $ \\
    11 &   7.90(0) &$  12.65(1)_{  -2.8\%}^{+  3.9\%} $ \\
    12 &   8.89(1) &$  14.31(1)_{  -2.7\%}^{+  3.6\%} $ \\
    13 &   9.89(1) &$  15.99(1)_{  -2.6\%}^{+  3.7\%} $ \\
    14 &  10.92(1) &$  17.72(1)_{  -2.5\%}^{+  3.5\%} $ \\
 \hline
 \end{tabular}
 \renewcommand{\baselinestretch}{1.0}
 \caption{Total cross sections for $ZZ$ production as a function of energy.
 The renormalisation scale and factorisation scales are $\mu_R=\mu_F=M_Z$.
 Vector bosons are produced exactly on-shell and no decays are included.}
 \label{ZZtot}
 \end{center}
 \end{table}
\renewcommand{\baselinestretch}{1.0}

The decay of a Higgs boson  to two $Z$'s, which subsequently decay to leptons, is a promising search
channel at the LHC. This is due to the fact that  the Higgs will decay to $Z$'s (with a moderate
branching ratio) over a large range of Higgs masses that are not presently excluded.  In addition,
the four lepton signature associated with $ZZ$ decay is experimentally clean. With Higgs searches in
mind we apply the following cuts,
\begin{eqnarray}
 p^{\ell_1,\ell_2}_{T} >  20~{\rm{GeV}}, \; p^{\ell_3,\ell_4}_{T} >  5~{\rm{GeV}},
 \; |\eta_{\ell}| < 2.5, \;\; m_{\ell\ell},m_{\ell'\ell'} > 5~{\rm{GeV}}  \;. \label{eq:offZZ} 
\end{eqnarray} 

\begin{figure}
\begin{center}
\includegraphics[width=12cm]{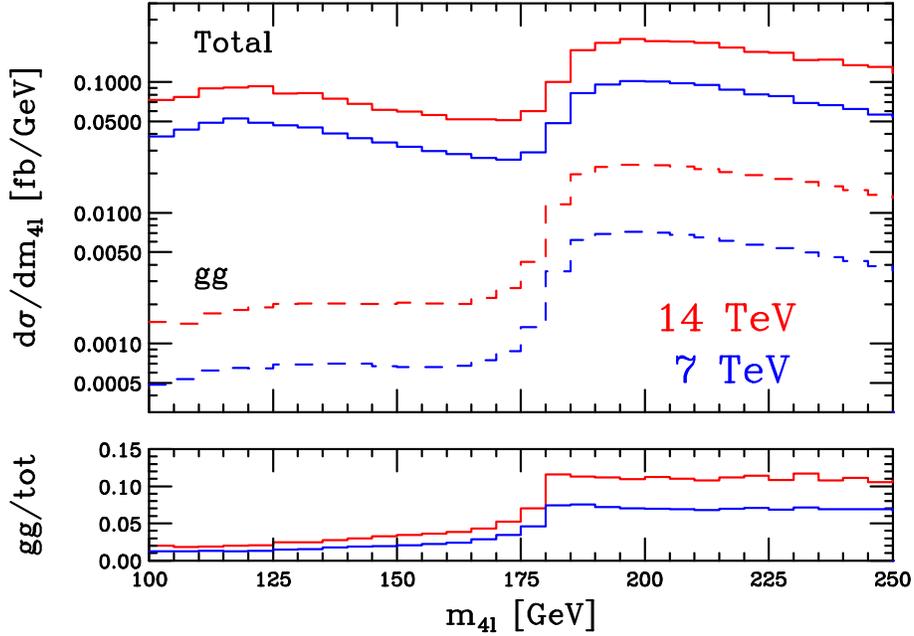}
\caption{The invariant mass of the four lepton system in $Z/\gamma^\star(\to e^+ e^-) Z/\gamma^\star(\to \mu^+ \mu^-)$
production at $\sqrt{s}=7$ and
$\sqrt{s}=14$ TeV, with the cuts of Eq.~(\protect\ref{eq:offZZ}). In the upper panel
we show both the total NLO prediction (upper curves) and the contribution from the $gg$ initial
state only (lower curves). In the lower panel we plot the fraction of the NLO prediction
resulting from the $gg$ initial state.
\label{fig:ZZm4l}}
\end{center}
\end{figure}

In this definition of the cuts, $\ell_1$ and $\ell_2$ represent the two hardest leptons and $\ell_3$
and $\ell_4$ represent the two sub-leading leptons. The relevant distribution for the Higgs
search is the invariant mass of the four-lepton system ($m_{4\ell}$), for which we present our
predictions in \fig~\ref{fig:ZZm4l}.
We show NLO predictions for both $\sqrt{s}=7$~TeV and $\sqrt{s}=14$~TeV, as well as the contribution
from the gluon-gluon diagrams alone.

From the figure we observe that, although the gluon initiated pieces are fairly important at the
level of the total cross section, their effect in the region $m_{4\ell} < 2 M_Z$ is rather smaller
(at the few percent level).  As this threshold is crossed the percentage effect increases to around
$7\%$~(7 TeV) or $10\%$~(14 TeV). Our results at $14$~TeV agree with the findings of a previous
study in a similar kinematic range~\cite{Binoth:2008pr}. It is clear that the $gg$ initiated piece is
most important as a background to Higgs bosons searches in the region $m_H > 2 M_Z$.


\section{Conclusions}
\label{sec:conc}

In this paper we have provided NLO predictions for all diboson processes
at the LHC, both at the current operating energy of $\sqrt{s}=7$~TeV and at
higher energies appropriate for future running. The calculations are contained
in the parton level code MCFM, which includes the 
implementation of $pp\rightarrow \gamma\gamma$ for the first time.
In addition, where appropriate we have revisited the treatment of many of the
vector boson pair processes in order to ensure the relevance of the
predictions for the LHC.

In order to enable simpler comparisons with experimental results we have implemented experimental photon
isolation cuts into MCFM.  This requires the inclusion of fragmentation
contributions~\cite{Catani:2002ny,Catani:1996vz}, in which a QCD parton fragments into a photon plus hadronic
energy. These fragmentation contributions require the introduction of fragmentation functions that contain both
non-perturbative and perturbative information.  Including this isolation condition extends the previous treatment
of photons in MCFM, for which the smooth cone isolation of Frixione~\cite{Frixione:1998jh} had been used.
Although this latter method is simple to implement theoretically it is not well-suited to experimental
studies.

At the LHC, contributions to diboson production which proceed through a gluon initiated quark loop can have a
significant effect on cross-sections. Although formally in perturbation theory they enter at NNLO the
large flux of gluons at LHC center of mass energies can overcome the formal ${\cal O}(\alpha_s^2)$ suppression.
Consequently we have included the gluon initiated processes $gg\rightarrow \{\gamma\gamma,Z\gamma, ZZ,WW\}$
whose contributions have been studied in the
past~\cite{Dicus:1987fk,Nadolsky:2007ba,Bern:2001df,Bern:2002jx,Ametller:1985di,vanderBij:1988fb,Adamson:2002rm,Dicus:1987dj,Glover:1988fe,Glover:1988rg,Binoth:2005ua,Binoth:2006mf,Matsuura:1991pj,Zecher:1994kb,:2008uu,Binoth:2008pr}. 
We have also included higher order corrections to the $gg\rightarrow\gamma\gamma$
process~\cite{Bern:2001df,Bern:2002jx}, which are formally at the level of N$^{3}$LO in perturbation theory.
These corrections are indeed of phenomenological relevance at the LHC, since at ``NLO" the gluon contribution
is around $20\%$ of the total cross section at $\sqrt{s}=7$~TeV.

We have presented detailed results for the diphoton process, $pp\rightarrow \gamma\gamma$ at the LHC, which
is an extremely important channel for light Higgs boson searches. We have
presented theoretical predictions for the $m_{\gamma\gamma}$ distribution using
experimental cuts and isolation. We have illustrated how these cuts reduce
the nominal cross section for a range of $\sqrt{s}$ appropriate to the
LHC and also investigated the sensitivity of our predictions to the
amount of hadronic energy in a fixed size isolation cone.
We have also shown that experimental Higgs search cuts, which usually require staggered photon
transverse momenta, produce large $K$-factors at NLO due to the limited kinematic configurations probed
at leading order.

We also presented results for $W\gamma$ and $Z\gamma$ production. As a result of the
new isolation procedures we were able to compare our NLO prediction for
the cross sections with the recently measured values from CMS.
We investigated the effects of various lepton cuts on cross-sections and
predictions for distributions, particularly those cuts designed to suppress
the contribution of photon radiation in the vector boson decay. This
is an important consideration in the search for anomalous couplings between
vector bosons and photons.
Although the gluon-gluon contribution to the cross section is a few percent of the total
at NLO, we found that as the minimum photon $p_{T} $ is increased the
gluon initiated terms become relatively more important.

We studied the effects of including gluon initiated processes on $WW$ and $ZZ$ production.  These pieces have been calculated using
various methods in the past~\cite{Binoth:2005ua,Binoth:2006mf,:2008uu,Binoth:2008pr}, where it has been shown that the effects of the
massive top quark are small. For this reason we have ignored the effect of a third
generation (for $WW$) or top quark loop (for $ZZ$) and instead only include loops of massless quarks using the analytic formulae
described in Appendix~\ref{app:amps}. For $ZZ$ production we also include the effects of singly-resonant diagrams that
had previously been neglected in ref.~\cite{Campbell:1999ah}. For the case of $WW$ production we paid particular attention to the
effects of a jet-veto on the NLO cross section, since at the LHC a jet-veto is necessary in order to reduce the abundant top
background. We found that applying a jet-veto and Higgs search cuts increases the overall percentage of the  cross section associated
with the gluon initiated process. For $ZZ$ production we found that the contribution of the $gg$ process is only important in
searches for a Higgs boson  with a mass greater than $2M_Z$. The fact that these $gg$ corrections can be large in some circumstances,
suggests that the two loop corrections are worth calculating, to get a better idea of the associated theoretical error.

We also presented results for $WZ$ production at the LHC. As an example, we investigated the fraction of events that 
survive a minimum cut on the $Z$ transverse momentum. This quantity is important in regards to boosted searches for Higgs bosons
and supersymmetry. 

\section*{Acknowledgements} 

We thank Joe Lykken for useful discussions and Adrian Signer for providing us with a copy
of the numerical program described in ref.~\cite{DeFlorian:2000sg}.  
Fermilab is operated by Fermi Research Alliance, LLC under
Contract No. DE-AC02-07CH11359 with the United States Department of Energy.

\appendix

\section{Input parameters for phenomenological results}
\label{app:inputs}

The electroweak parameters that we regard as inputs are,
\begin{eqnarray}
M_W &=& 80.398~\mbox{GeV}\;, \;\; M_Z=91.1876~\mbox{GeV}\;, \\
\Gamma_W &=& 2.1054~\mbox{GeV}\;, \;\; \Gamma_Z=2.4952~\mbox{GeV} \;, \\
G_F &=& 1.16639 \times 10^{-5} \, \mbox{GeV}^{-2}\;.
\end{eqnarray}
Using the values of $M_W$, $M_Z$ and $G_F$ as above then determines 
$\alpha_{e.m.}(M_Z)$ and $\sin^2\theta_w$ as outputs, where $\theta_w$ is
the Weinberg angle. We find,
\begin{eqnarray}
\sin^2\theta_w &=& 1 - M_W^2/M_Z^2 = 0.222646 \;, \\
\alpha_{e.m.}(M_Z) &=& \frac{\sqrt2 G_F M_W^2 \sin^2\theta_w}{\pi}
 = \frac{1}{132.338} \;.
\end{eqnarray}
This value of $\alpha_{e.m.}$ may not correspond to the value of $\alpha_{e.m.}$ used 
to fit the fragmentation functions in ref.~~\cite{Bourhis:1997yu}. The value of $\alpha_{e.m.}$
used in their fit is hard to extract from ref.~~\cite{Bourhis:1997yu}. After isolation we believe that
any potential mismatch will be of minor numerical significance.

\section{Helicity amplitudes for gluon-gluon processes}
\label{app:amps}

In this appendix we present results for three of the gluon initiated
processes considered in the text, namely $gg \to Z\gamma$, $gg \to WW$
and $gg \to ZZ$. We first describe some general notation and then
consider each of these processes in turn.


\subsection{Notation}
In order to specify the amplitudes we first introduce some notation.
The QED and QCD couplings are denoted by $e$ and $g_s$ respectively
and $Q^q$ is the charge of quark $q$ in units of $e$. The ratio of
vector boson $V$ ($=W,Z$) and photon propagators is given by,
\begin{equation}
\prop{V}(s) = {s \over s - M_V^2 + i \,\Gamma_V \, M_V}\,,
\end{equation}
where $M_V$ and $\Gamma_V$ are the mass and width of the boson $V$.  
Fermions interact with the $Z$ boson through the following
left- and right-handed couplings,
\begin{eqnarray}
v_L^e & = & { -1 + 2\sin^2 \theta_w \over \sin 2 \theta_w } \;, 
\hskip 2.3 cm 
v_R^e  = { 2 \sin^2 \theta_w \over  \sin 2 \theta_w } \;,  \cr 
v_L^q & = & { \pm 1 - 2 Q^q\sin^2 \theta_w \over  \sin 2 \theta_w } \;,
\hskip 1.9 cm 
v_R^q = -{ 2 Q^q \sin^2 \theta_w \over \sin 2 \theta_w }  \;.
\end{eqnarray}
The subscripts $L$ and $R$ refer to whether the particle to which 
the $Z$ couples is left- or right-handed and the
two signs in $v_L^q$ correspond to up $(+)$ and down $(-)$ type quarks

We express the amplitudes in terms of spinor products defined as,
\begin{equation}
\spa i.j=\bar{u}_-(p_i) u_+(p_j), \;\;\;
\spb i.j=\bar{u}_+(p_i) u_-(p_j), \;\;\;
\spa i.j \spb j.i = 2 p_i \cdot p_j\,.
\end{equation}

\subsection{Amplitudes for $gg \to Z \gamma$}
\begin{figure}
\begin{center}
\includegraphics[angle=270,scale=0.5]{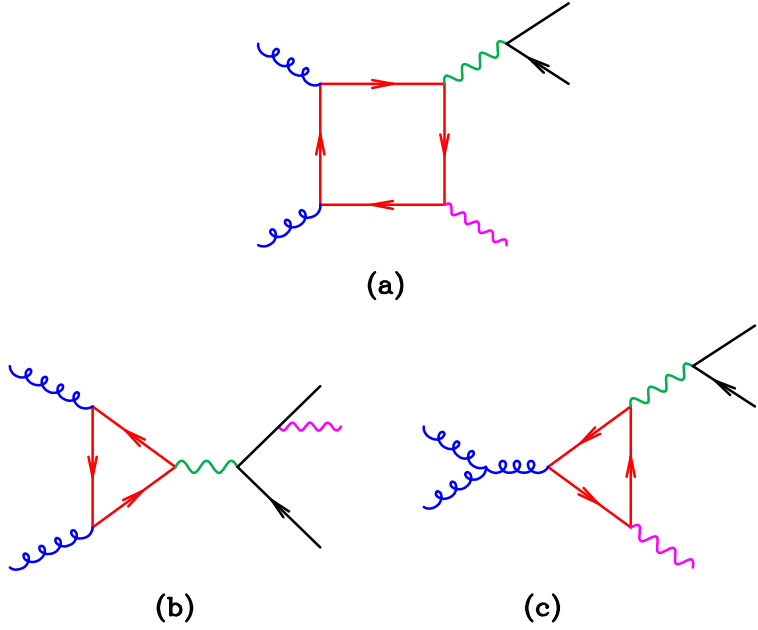}
\end{center}
\caption{Examples of diagrams that could potentially contribute to
$gg \to \gamma \ell \overline{\ell}$.}
\label{Zgam}
\end{figure}

In this section we present results for the amplitudes relevant for the process,
\begin{equation}
 0 \rightarrow g(p_1) + g(p_2) + \gamma(p_3) + \ell(p_4) + \overline{\ell}(p_5) \;.
\end{equation}
These amplitudes can be extracted from the fermion loop amplitudes 
$A_6^{v,ax}(1_q,2_\qb,3_g,4_g,5_{\overline{\ell}},6_{\ell})$ in ref.~\cite{Bern:1997sc},
by taking the limit in which the quark and antiquark are collinear, and renaming the momenta.
The boxes and triangles which potentially could contribute to this amplitude are shown in \fig~\ref{Zgam}.
Considering the triangle diagrams first, we see that diagrams like 
\fig~\ref{Zgam}(c) can never give a contribution because of colour.
\fig~\ref{Zgam}(b) with the vector coupling of the $Z$ to the fermion loop 
vanishes because of Furry's theorem, and with the axial coupling to the 
fermion loop vanishes because of Bose statistics (Landau-Yang theorem).
We therefore only have to consider box diagrams like \fig~\ref{Zgam}(a).
In this case the diagrams with an axial coupling vanish, so we only have to
consider box diagrams with a vector coupling.

The result for the fully dressed amplitude is,
\begin{eqnarray}
&&{\cal A}_{5}^{1\rm -loop}(1_g,2_g,3_{\gamma},4_{\ell},5_{\overline{\ell}})  
= 2 \sqrt{2} e^3 \frac{g_s^2}{16 \pi^2} \delta^{a_1 a_2} \nn \\ 
&\times & \sum_{i=1}^{\nf} Q^i \left[ -Q^i + {1\over 2} v_{L,R}^e 
             (v_L^i+v_R^i) \prop{Z}(s_{56}) \right] A^v(1_g,2_g,3_{\gamma},4_{\ell},5_{\overline{\ell}}) \;,
\end{eqnarray}
where $a_1,a_2$ are the colour labels of the two gluons and there are $n_f$ flavours of massless quarks
circulating in the loops.



The amplitude 
$A^{(v)}_5(1_g^+,2_g^+,3_{\gamma}^+,4_{\ell}^-,5_{\overline{\ell}}^+)$ is 
entirely rational and given by the following expression,
\begin{equation}
A^{v}_5(1_g^+,2_g^+,3_{\gamma}^+,4_{\ell}^-,5_{\overline{\ell}}^+)=
 2 \Bigg\{ \bigg[\frac{ \spa1.4^2 \spb3.1}{ \spa1.2^2 \spa1.3 \spa4.5}
 -\frac{1}{2}\frac{ \spb5.3^2}{ \spa1.2^2 \spb5.4}\bigg]
  + \bigg[1\leftrightarrow2\bigg] \Bigg\} \;.
\end{equation}
The amplitude 
$A^{(v)}_5(1_g^+,2_g^+,3_{\gamma}^-,4_{\ell}^-,5_{\overline{\ell}}^+)$ 
contains dependence on the box and triangle functions $\Ll_0(r)$, $\Ll_1(r)$ and
$\Ls_{-1}(r_1,r_2)$ that will be defined below. The result is,
\begin{eqnarray}
  A^{(v)}_5(1_g^+,2_g^+,3_{\gamma}^-,4_{\ell}^-,5_{\overline{\ell}}^+)
 &=&  +2\Bigg\{\frac{\spa1.3^2 \spa2.4^2+\spa1.4^2 \spa2.3^2}{ \spa1.2^4 \spa4.5} 
{\rm{Ls}}_{-1}\left(\frac{-s_{13}}{-s_{123}},\frac{-s_{23}}{-s_{123}}\right) \nonumber\\
&+&\bigg[2\frac{ \spa2.3 \spa1.4 \spa2.4 \spb2.1}{\spb3.1 \spa1.2^3 \spa4.5} {\rm{L}}_0\left(\frac{-s_{123}}{-s_{13}}\right)
 -\frac{ \spa2.4^2 \spb2.1^2}{\spb3.1^2 \spa1.2^2 \spa4.5} {\rm{L}}_1\left(\frac{-s_{123}}{-s_{13}}\right) \nonumber\\ 
&-&\frac{ \spa1.3 \spa2.4 \spb2.1 \spb5.1}{ \spb3.1 \spa1.2^2 \spa4.5 \spb5.4 }\bigg]
 + \bigg[1\leftrightarrow2\bigg] \Bigg\} \;.
\label{++-}
\end{eqnarray}
We note that the $1 \leftrightarrow 2$ symmetry in this equation is to be applied to the terms
inside square brackets only.

The final helicity amplitude 
$A^{(v)}_5(1_g^+,2_g^-,3_{\gamma}^-,4_{\ell}^-,5_{\overline{\ell}}^+)$ can be obtained by exchange from Eq.~(\ref{++-}),
\begin{eqnarray}
A^{(v)}_5(1^+,2^-,3_{\gamma}^-,4_{\ell}^-,5_{\overline{\ell}}^+)&=&
-2\Bigg\{\frac{ \spb3.1^2 \spb2.5^2+ \spb3.5^2 \spb2.1^2 }{ \spb3.2^4 \spb5.4} 
{\rm{Ls}}_{-1}\left(\frac{-s_{13}}{-s_{123}},\frac{-s_{12}}{-s_{123}}\right) \nonumber\\
&+&\bigg[2\frac{ \spb2.1 \spb3.5 \spb2.5 \spa2.3}{\spa1.3 \spb3.2^3 \spb5.4} {\rm{L}}_0\left(\frac{-s_{123}}{-s_{13}}\right)
-\frac{ \spb2.5^2 \spa2.3^2}{\spa1.3^2 \spb3.2^2 \spb5.4} {\rm{L}}_1\left(\frac{-s_{123}}{-s_{13}}\right) \nonumber\\ 
&-&\frac{ \spb3.1 \spb2.5 \spa2.3 \spa4.3}{\spa1.3 \spb3.2^2 \spb5.4 \spa4.5}\bigg]
 + \bigg[2\leftrightarrow3\bigg] \Bigg\} \;,
\end{eqnarray}
where again the $2 \leftrightarrow 3$ is to be applied to the terms inside square brackets only.

The latter two amplitudes are defined in terms of the following functions that arise from 
box integrals with one non-lightlike external line,
\begin{eqnarray}
  \Ll_0(r) &=& {\ln(r)\over 1-r} \;, \nn \\
  \Ll_1(r) &=& {\Ll_0(r)+1\over 1-r} \;, \nn \\
  \Ls_{-1}(r_1,r_2) &=& \Li_2(1-r_1) + \Li_2(1-r_2) + \ln r_1\,\ln r_2 - {\pi^2\over6} \;,
\end{eqnarray}
and the dilogarithm is defined by,
\begin{equation}
\Li_2(x) = - \int_0^x dy \, {\ln(1-y) \over y}\ \;.
\end{equation}
The remaining amplitudes can be obtained from these ones by simple symmetry operations.


\subsection{Amplitudes for $gg \to WW$}
In this section we present results for the amplitudes relevant for the process,
\begin{equation}
 0 \rightarrow g(p_1) + g(p_2) + \nu_\ell(p_3) + \overline{\ell}(p_4) + \ell^\prime(p_5) + \overline{\nu}_{\ell^\prime}(p_6) \;.
\end{equation}
\begin{figure}
\begin{center}
\includegraphics[angle=270,scale=0.5]{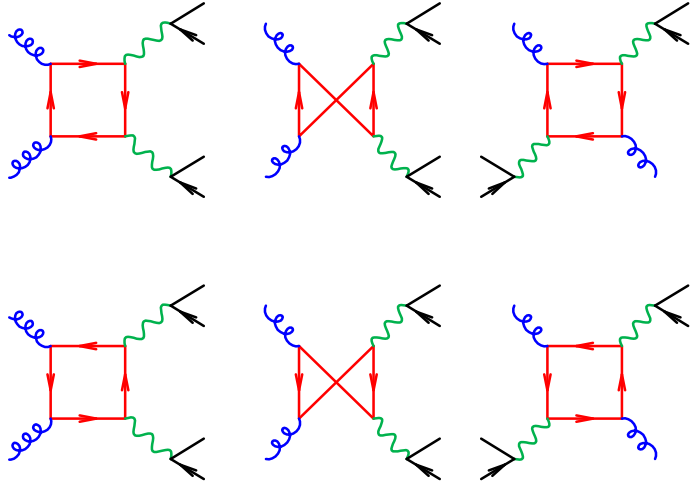}
\end{center}
\caption{Diagrams that contribute to $gg \to WW$.}
\label{WWfig}
\end{figure}
There are six contributing Feynman diagrams depicted in \fig~\ref{WWfig}. 
These diagrams represent exactly the same set that appears
in the calculation of certain contributions to the process $e^+ e^- \to \mbox{4 partons}$. We therefore
simply reinterpret the compact expressions for such amplitudes presented in ref.~\cite{Bern:1997sc},
modifying the overall factor appropriately.
Specifically, we find that the contribution from a single generation of
massless quarks in the loop is given by,
\begin{eqnarray}
&& {\cal A}_{6}^{1\rm -loop}\left(1_g^{h_1},2_g^{h_2},3_{\nu_\ell}^-,4_{\overline{\ell}}^+,5_{\ell^\prime}^-,6_{\overline{\nu}_{\ell^\prime}}^+ \right)
\nn \\
&=& \delta^{a_1 a_2} \left(\frac{g_w^4 g_s^2}{16\pi^2}\right) \prop{W}(s_{34}) \prop{W}(s_{56})
 \, \left[ A_{6;4}^v \left(3_q^+, 4_{\bar q}^-, 1_g^{-h_1}, 2_g^{-h_2}; 6_{\overline e}^-, 5_e^+ \right) \right]^{cc} \;.
\end{eqnarray}
The helicities and colour labels of the two gluons are $h_1,h_2$ and $a_1,a_2$ respectively
and the amplitude $A_{6;4}^v$ is defined in Sections 2 and 11 of ref.~\cite{Bern:1997sc}.
The operation $[ \ldots ]^{cc}$ exchanges the spinor products $\spa a.b$ and $\spb a.b$ in the
amplitude.
The labelling on the right hand side of this equation is as written in ref.~\cite{Bern:1997sc}. For our purposes we make the
identification on the left hand side, ($q \to \nu_\ell$, $\bar q \to \overline{\ell}$, $e \to \ell^\prime$ and
$\bar e \to {\overline{\nu}_{\ell^\prime}}$).


\subsection{Amplitudes for $gg \to ZZ$}
In this section we present results for the amplitudes relevant for the process,
\begin{equation}
 0 \rightarrow g(p_1) + g(p_2) + \ell(p_3) + \overline{\ell}(p_4) + \ell^\prime(p_5) + \overline{\ell^\prime}(p_6) \;.
\end{equation}
The extension of the procedure outlined above for $gg \to WW$ is clear. One must now simply sum over all four possible
helicity combinations for the leptonic decays.
The result for the fully dressed amplitude is, for a particular choice of lepton helicities,
\begin{eqnarray}
&&{\cal A}_{6}^{1\rm -loop}\left(1_g^{h_1},2_g^{h_2},3_{\ell}^-,4_{\overline{\ell}}^+,5_{\ell^\prime}^-,6_{\overline{\ell^\prime}}^+ \right)
\nn \\
&=&  \delta^{a_1 a_2} \, \frac{e^4 g_s^2}{2 \pi^2}
 \sum_{i=1}^{\nf} \left[
     \left( -Q^i + {1\over 2} v_{L,R}^\ell (v_L^i+v_R^i) \prop{Z}(s_{34}) \right)
     \left( -Q^i + {1\over 2} v_{L,R}^{\ell^\prime} (v_L^i+v_R^i) \prop{Z}(s_{56}) \right) \right. \nn \\
&& \left.
    + {1\over 4} v_{L,R}^\ell v_{L,R}^{\ell^\prime} {(v_L^i-v_R^i)}^2 \, \prop{Z}(s_{34}) \prop{Z}(s_{56}) \right]
 \, \left[ A_{6;4}^v \left(3_q^+, 4_{\bar q}^-, 1_g^{-h_1}, 2_g^{-h_2}; 6_{\overline e}^-, 5_e^+ \right) \right]^{cc} .
\end{eqnarray}
The labelling on the right hand side of this equation is as written in ref.~\cite{Bern:1997sc}. For our purposes we make the
identification on the left hand side, ($q \to \ell$, $\bar q \to \overline{\ell}$, $e \to \ell^\prime$ and $\bar e \to \overline{\ell^\prime}$).

\newpage
\bibliography{CEW}
\bibliographystyle{JHEP}

\end{document}